\documentclass{aa}  

\usepackage{lipsum}
\usepackage{hyperref}
\usepackage{txfonts}
\usepackage{graphicx}	% Including figure files
\usepackage{amsmath}	% Advanced maths commands
\usepackage{amssymb}	% Extra maths symbols
\usepackage{subcaption}
\usepackage{float}
\usepackage{enumerate}
\usepackage{multirow}
\usepackage[export]{adjustbox}
\usepackage{tabularx}

\begin{document}

   \title{TP-AGB stars and stellar population properties of a post-starburst galaxy at $z\sim2$ through optical and NIR spectroscopy with JWST}
   \titlerunning{JADES 138717}

   \author{Davide Bevacqua\inst{1,2,10}\thanks{E-mail: davide.bevacqua@inaf.it}
             \and
          Paolo Saracco\inst{1}
          \and
          Francesco La Barbera\inst{3}
          \and
          Guido De Marchi\inst{4}
          \and
          Roberto De Propris\inst{5, 6}
          \and
          Fabio R. Ditrani\inst{7, 1}
          \and
          Anna R. Gallazzi\inst{8}
          \and
          Giovanna Giardino\inst{9}
          \and
          Danilo Marchesini\inst{10}
          \and
          Anna Pasquali\inst{11}
          \and
          Tim D. Rawle\inst{12}
          \and
          Chiara Spiniello\inst{13, 3}
          \and 
          Alexandre Vazdekis\inst{14, 15}
          \and
          Stefano Zibetti\inst{8}
          }

   \institute{INAF - Osservatorio Astronomico di Brera, via Brera 28, 20121 Milano, Italy              
         \and
             DiSAT, Universit\'{a} degli Studi dell'Insubria, via Valleggio 11, I-22100 Como, Italy
             \and
             INAF - Osservatorio Astronomico di Capodimonte, Via Moiariello 16, 80131, Naples, Italy
             \and
             European Space Research and Technology Centre, Keplerlaan 1, 2200 AG Noordwijk, Netherlands
             \and
             FINCA, University of Turku, Vesilinnantie 5, Turku, 20014, Finland
             \and
             Department of Physics and Astronomy, Botswana International University of Science and Technology, Private Bag 16, Palapye, Botswana
             \and
             Universit\'{a} degli studi di Milano - Bicocca, Piazza della scienza, I-20125 Milano, Italy
             \and
             INAF - Osservatorio Astronomico di Arcetri, Largo Enrico Fermi 5, I-50125, Firenze, Italy   
             \and
             ATG Europe for the European Space Agency, European Space Research and Technology Centre, Noordwijk, Netherlands
             \and
             Physics and Astronomy Department, Tufts University, 574 Boston Avenue, Medford, MA 02155, USA
             \and
             Astronomisches Rechen-Institut, Zentrum f\"{u}r Astronomie der Universit{\"a}t Heidelberg, M\"{o}nchhofstrasse 12 - 14, 69120 Heidelberg, Germany
             \and
             European Space Agency (ESA), European Space Astronomy Centre (ESAC), Camino Bajo del Castillo s/n, 28692 Villanueva de la Ca\~{n}ada, Madrid, Spain
             \and
             Sub-Dep. of Astrophysics, Dep. of Physics, University of Oxford, Denys Wilkinson Building, Keble Road, Oxford OX1 3RH, UK
             \and
             Instituto de Astrof\'{i}sica de Canarias (IAC), La Laguna, E-38200 Tenerife, Spain
             \and
             Departamento de Astrof\'{i}sica, Universidad de La Laguna, E-38205 Tenerife, Spain
           }

   \date{Received 13 January 2025 / Accepted 21 May 2025}
 
  \abstract
  {We present a detailed optical and NIR spectral analysis of J-138717, a post-starburst galaxy at $z = 1.8845$ observed with JWST/NIRSpec, for which we derive a stellar mass of $3.5 \pm 0.2 \times 10^{10}$ M$_\odot$ and a stellar velocity dispersion of $198 \pm 10$ km s$^{-1}$. We estimate an age of $\sim0.9$ Gyr and a sub-solar metallicity (between $-0.4$ and $-0.2$ dex). We find generally consistent results when fitting the optical and NIR wavelength ranges separately or using different model libraries. The reconstruction of the star formation history indicates that the galaxy assembled most of its mass quickly and then rapidly quenched, $\sim0.4$ Gyr prior to observation. Line diagnostics suggest that the weak emission is probably powered by residual star formation (SFR$\sim0.2$M$_\odot$ yr$^{-1}$) or a low-luminosity AGN, with no strong evidence for outflows in ionized or neutral gas. We perform a detailed study of the NIR spectral indices by comparing observations with predictions of several state-of-the-art stellar population models. This is unprecedented at such a high redshift. In particular, the analysis of several CO and CN features argues against a heavy contribution of Thermally Pulsating (TP-)AGB stars. Observations align better with models that include a minimal contribution from TP-AGB stars, but they are also consistent with a mild contribution from TP-AGB stars, assuming a younger age (consistent with the fits). The analysis of other NIR spectral indices shows that current models struggle to reproduce observations. This highlights the need for improved stellar population models in the NIR, especially at young ages and low metallicities, which is most relevant for studying high redshift galaxies in the JWST era.}


   \maketitle
%
%-------------------------------------------------------------------
\nolinenumbers

\section{Introduction}\label{sect:intro}

One of the most debated topics of modern astrophysics is how massive quiescent galaxies form and quench. Both observations and simulations suggest that quiescent galaxies can undergo different formation paths, implying a variety of star formation histories (SFHs; e.g., \citealt{Bevacqua+24}) and quenching mechanisms (e.g., \citealt{Tacchella+22}), depending on the properties of the gas from which they formed and the subsequent mass growth and assembly, which can vary significantly depending on the local cosmological conditions and environment (e.g., \citealt{DeLucia+06}).

Concerning their quenching, many processes can be responsible for the halt of star formation in galaxies (e.g., \citealt{ManBelli18} and references therein) and there is no current agreement on what the dominant quenching mechanism is, despite the efforts of the past decades \citep[e.g., ][]{Thomas+05, McDermid+15, Newman+18, Rowlands+18, Belli+19, Bluck+22, Piotrowska+22, deLucia+24}. Quenching mechanisms can be broadly distinguished into rapid (< 1 Gyr), like ejective AGN feedback and ram-pressure stripping, and slow (> 1 Gyr), like virial shocks and thermal AGN feedback. Both simulations (e.g., \citealt{Rodriguez-Gomez+19}) and observations (e.g., \citealt{Tacchella+22}) find a broad range of quenching timescales for galaxies in the local Universe, implying that nearby quiescent galaxies experienced different quenching mechanisms. Alternatively, galaxies can simply run out of gas as a consequence of highly efficient star formation (e.g. \citealt{Vazdekis+96, Vazdekis+97, SR+20})%. In this case, some residual, low-level star formation is expected to persist (e.g., \citealt{SR+20}).

At high redshifts, the processes responsible for the galaxy quenching are typically shorter, since the Universe is younger. For instance, massive quiescent galaxies found at $z\gtrsim3$ (e.g., \citealt{Forrest+20_a, Forrest+20_b, Saracco+20, Glazebrook+24}) require very high star formation efficiencies and short quenching times. As a consequence of an efficient starburst followed by rapid quenching, it is expected that high-redshift quiescent galaxies resemble local post-starburst galaxies (PSBs), namely very young galaxies with negligible ongoing star formation. Observations seem to support this. For instance, using a sample of 9 galaxies at $z\sim3$, \citet{DEugenio+20} showed that the typical near-UV/optical spectrum of high-$z$ quiescent galaxies is characterized by a strong Balmer break, larger than the 4000 \AA \, break, and corresponding to ages of a few hundreds of Myr, typical of PSBs. The ignition of the starburst and the main cause of the rapid quenching of these galaxies are though still unclear.

So far, due to instrumental limitations, only a limited number of (massive) quiescent galaxies have been observed at high redshifts. The advent of the James Webb Space Telescope (JWST) has opened the possibility of performing large galaxy surveys at high redshift, allowing us to study the properties of a large number of high-redshift quiescent galaxies at different masses. Recent observations with JWST have revealed the presence of a large number of quiescent galaxies at very high redshifts, up to $z=7.3$ \citep{Looser+24, Weibel+24}, spanning a wide range of masses \citep{gs9209, Labbe+23, Looser+23, Looser+24, rubies, Slob+24, degraaff_z5, Bingjie+24, Carnall+24, Glazebrook+24, Alberts+24}. Several low-mass ($<10^{10}$ M$_\odot$) quiescent galaxies at high redshifts exhibit bursty SFHs, i.e. they are characterized by episodic events of short star formation followed by periods of prolonged quiescence, suggesting that they are just temporary (or `mini'-) quenched \citep{Looser+23, Looser+24, Strait+23, Dome+24}. For these galaxies, quenching via environmental interactions or ejective stellar feedback is currently favored against other mechanisms, such as supernovae feedback \citep{Looser+23, Sandles+23, Strait+23, Gelli+23, Gelli+24, Carniani+24, Zhang+24, Cutler+24, Hamadouche+24}. For massive galaxies, many studies have reported a high incidence of strong AGN activity in recently quenched galaxies with PSB-like spectra associated with massive, high-velocity multiphase gas outflows (\citealt{Belli+24, Davies+24, Park+24, Scholtz+24, Wu24, DEugenio_psb}; see also \citealt{Tremonti+07, Baron+17, FS+19, Maltby+19}), suggesting that ejective AGN feedback plays a major role in quenching massive galaxies at high redshifts, in agreement with some simulations \citep{Pontzen+17, Davis+19, Zheng+20}. However, the population of high-redshift quiescent galaxies generally exhibit a rich variety of SFHs (e.g., \citealt{Carnall+24, Park+24, Nanayakkara+25}), in agreement with studies at lower redshifts (e.g., \citealt{Thomas+05, Panter+07, McDermid+15, Pacifici+16, Tacchella+22, Bevacqua+24}).

A major issue related to the estimate of the properties of these high-redshift galaxies, including their mass and SFH, is how different stellar population models affect the inferred properties. It is well known that stellar population models struggle to reproduce the near-infrared (NIR) spectrum of galaxies (e.g., \citealt{Eminian+08, Riffel+15, Riffel+19, eftekhari+21}). For quiescent galaxies, the characterization of the NIR spectrum is crucial, as a significant fraction of the stellar light is emitted in that wavelength region. Over the last decade, a lot of efforts have been put in this regard, and state-of-art models can now rely on better and more complete stellar libraries, as well as on the characterization of reliable NIR spectral indices \citep{Riffel+19, Francois+19, morelli+20, eftekhari+21}. Despite the improvements, the (few) studies based on local galaxies have shown that models still struggle to reproduce the NIR spectral indices (e.g., \citealt{eftekhari_CO, FLB+24}). 

One of the major sources of uncertainty for the characterization of the NIR spectrum of galaxies is the contribution of thermally pulsating (TP-)AGB stars. TP-AGB stars are intermediate-mass stars ($1-10$M$_\odot$) whose luminosity is maintained by a double-shell burning regime determining the pulsation. From the fuel-consumption theorem \citep{RV81}, it is estimated that TP-AGB stars dominate the galaxy spectrum at ages between 0.2 and 2 Gyr, peaking at about 1 Gyr, when they are expected to contribute up to $\sim80\%$ in the K-band and up to $\sim 40\%$ to the bolometric luminosity \citep{M98, M05, Marigo+08}. Including them in the models' recipes then has a significant impact on the inferred galaxies' properties \citep{Pastorelli+20}. For example, \citet{M06} showed that models including a strong contribution of TP-AGB predict systematically lower ages and masses, up to $\sim 60\%$, compared to models having a low contribution.

After decades, the disagreement persists due to several reasons. Theoretically, the modeling of this evolutionary phase is hampered by our poor understanding of many key processes of stellar evolution (e.g., mass-loss, dredge-up, varying opacities, etc.). As a result, different models adopting different recipes lead to different results. Observationally, TP-AGB stars are rare objects, because of their short lifetime, and only small samples of TP-AGB stars are observed in a few local globular clusters and the Magellanic clouds. Further, it has been shown that estimates of the TP-AGB contribution to the integrated light of galaxies derived from these samples are biased toward large values \citep{Girardi+13}. Additionally, TP-AGB stars are expected to give a small contribution to the optical spectrum of galaxies, whereas they are prominent in the NIR where observations, before the JWST era, have been severely limited.

Ideally, the best galaxy candidates to probe the contribution of TP-AGB stars are high-redshift PSBs, as they are young systems with no ongoing star formation and no strong contamination from cosmological evolution. However, observations have provided, so far, conflicting results, leading to a vivid debate on the actual contribution of TP-AGB stars. For instance, \citet{Kriek+10} found a low contribution of TP-AGB stars in PSBs observed at $0.7 < z < 2$, while \citet{Capozzi+16} found a heavy contribution for a sample of passive galaxies at $1.3 < z < 2.7$, in agreement with recent results by \citet{Lu+24} using prism observations with JWST of three PSBs. At lower redshifts, \citet{CG10} found that models with a low contribution of TP-AGB stars reproduce well the NIR colors of nearby PSBs, while models with a heavy contribution do not. This result was further corroborated by \citet{Zibetti+13} using spectroscopic data. On the other hand, \citet{MacArthur+10} showed that the bulges of local spiral galaxies require a heavy contribution. In between these results, \citet{Melnick+14} showed that models with a moderate contribution of TP-AGB stars can also reproduce the NIR photometry of local PSBs. In brief, the results of the literature are mixed.

One problem with previous results in the literature is that they are based on broad-band photometry or optical spectroscopy or low-resolution NIR spectra, while good-quality NIR spectroscopic data are required for an accurate investigation of the spectral features. For instance, \citet{Riffel+15} showed that, although nearby galaxies can exhibit strong features characteristic of TP-AGB stars (e.g., CO and CN molecular bands), well-calibrated models based on empirical stellar libraries perform better in reproducing the observations than models with a significant contribution of TP-AGB stars (see also \citealt{DH+18}). More recently, \citet{eftekhari_CO} showed that there is a large mismatch between the observed CO indices in local massive ETGs and the predictions of state-of-art stellar population models, although local massive quiescent galaxies might not be an optimal sample to probe the TP-AGB phase, as they are very old while the contribution of TP-AGB stars peaks at young ages. Thanks to JWST, it is now possible to obtain spectra of high-redshift, young quiescent galaxies with sufficiently high resolution to probe the contribution of TP-AGB stars directly from the spectral indices. 

In this work, we perform a spectral analysis of a quiescent PSB observed at $z\sim2$. In section \ref{sect:datatemp} we describe the JWST data and libraries used to analyze it. In section \ref{sect:fits} we estimate the physical and stellar population properties and reconstruct the SFH of the galaxy from the rest-frame optical and IR spectra. In section \ref{sect:indices} we perform a detailed analysis of the optical and NIR spectral indices. In Sect. \ref{sect:discussion} we discuss our results and summarize them in section \ref{sect:summary}. In this paper, we adopt a flat $\Lambda$CDM cosmology with H$_0 = 70$ km s$^{-1}$ and $\Omega_{\rm m} = 0.3$.

\section{Data and templates}\label{sect:datatemp}

\subsection{Observations and data reduction issues}\label{sect:data}

\begin{figure*}
\includegraphics[width=\textwidth]{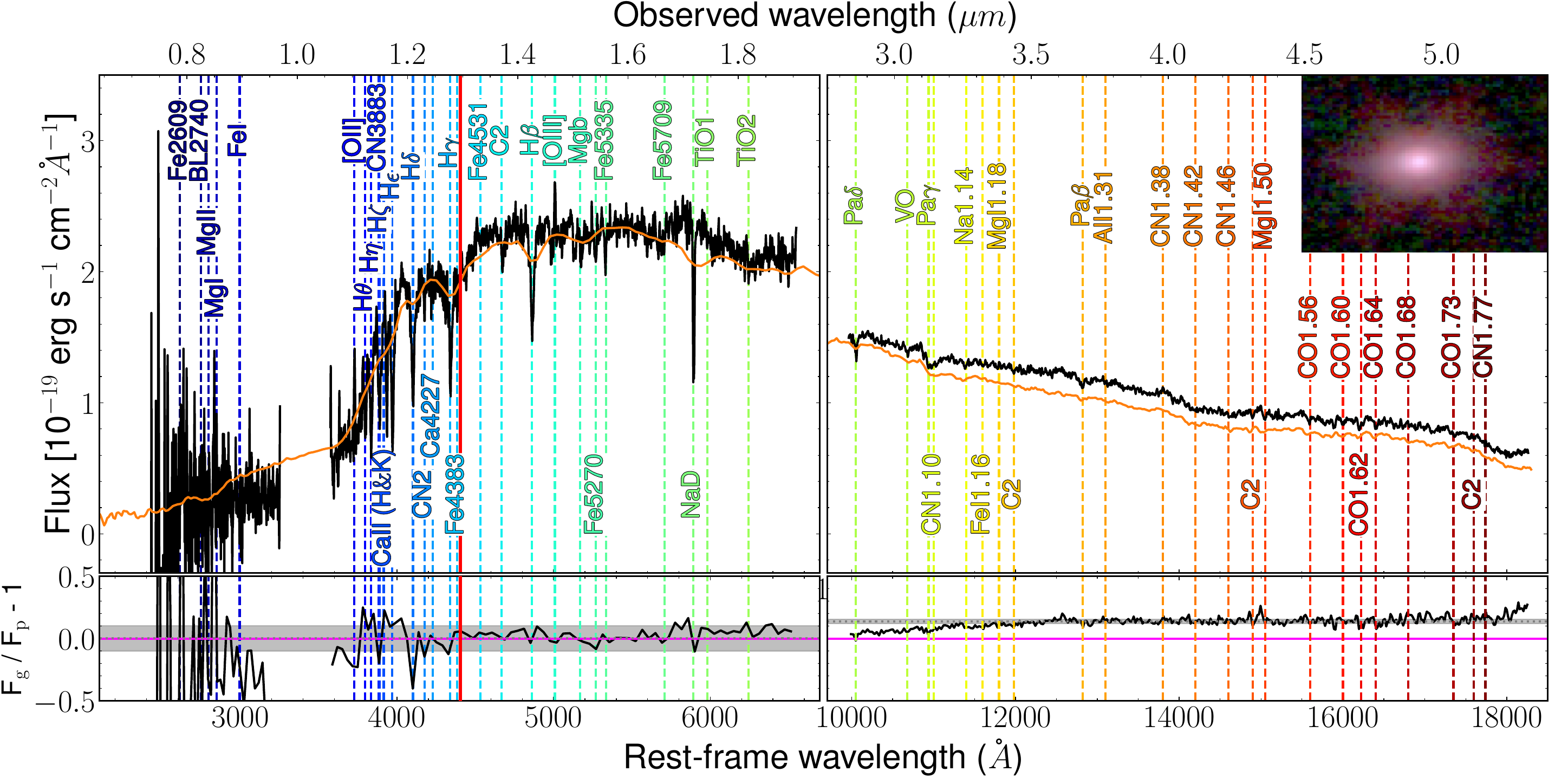}
\caption{The JWST spectrum of J-138717. The rest-frame wavelengths are calculated with respect to the spectroscopic redshift, z = 1.8845 (see sect. \ref{sect:kinematics}). The combined RGB imaging of the galaxy is shown in the upper right corner. The observed spectra are shown in the upper panels. The black lines are the medium resolution G140M/F070LP and G395M/F290LP spectra, while the orange line overplotted is the prism spectrum (we note that the prism spectrum also covers the rest-frame interval $\sim 6500 - 10000 \, \AA$ ; see e.g. Fig. \ref{fig:dr3_b24}). The two lower panels show the comparison of the flux of the prism and grating spectra (black line). In both panels, the magenta line marks a difference of 0, while the gray lines and shadow regions indicate the median and the average errors, respectively. The red vertical line in the left panels marks $\lambda = 1.27 \mu m$, above which the spectrum is potentially affected by second-order contamination. In all panels, the vertical dashed lines correspond to different absorption and emission lines. }
\label{fig:spec}
\end{figure*}

We selected our target galaxy from the JWST Advanced Deep Extragalactic Survey (JADES) \citep{JADES, jades_bunker}. For the spectroscopic data, we considered the data reduction of the third data release (\citealt{DR3}, hereafter DR3) of JADES. The spectroscopic data consists of observations with the two medium resolution (R $\sim 1000$) gratings, G140M/F070LP and G395M/F290LP. For the photometric data, we considered the public JADES catalog \citep{jades_photocat}. To select our target, we initially considered only galaxies having spectroscopic observations with both the prism and gratings observed at $z>1$. Then, we used the EAZY code \citep{eazy} to fit the photometry and estimate the rest-frame U$-$V and V$-$J colors, from which we selected quiescent galaxies using eq. 4 of \citet{UVJ}. Finally, we considered galaxies having SNR $>15$ in all spectra. The final sample consists of two galaxies having IDs 138717 (hereafter, J-138717) and 199773. In this paper, we only study the former; the latter is presented in \citet{DR3}. The photometric data of J-138717 consist of three NIRCam observations with filters F182M, F210M, and F444W, and four HST observations with filters F606W, F775W, F814W, F850LP. For this work, we considered the Kron photometry convolved by the point-spread function to the F444W spatial resolution (see \citealt{jades_photocat} for details). The photometric catalog also provides the photometric redshift, $z_{\rm phot} = 2.16$, and the half-light radius, $\approx 0.13$ arcsec, corresponding to about 1.1 kpc. The total integration time for the G140M, G395M, and the prism spectra are 9.2 hr, 35.6 hr, and 40.5 hr, resulting in a median signal-to-noise ratio (SNR) of about $19 \, \AA^{-1}$, $54 \, \AA^{-1}$, and $59 \, \AA^{-1}$, respectively. We show the two grating spectra, the prism spectrum, and the composite imaging of J-138717 in Figure \ref{fig:spec}. 

In Sect. 5 of \citet{DR3}, the authors show that the JADES spectra suffer from flux calibration issues. In the example shown in their Fig. 5, they report a 10-15$\%$ mismatch in the flux for G140M, with respect to the prism spectrum, at wavelengths above $\sim 1.3 \, \mu m$. However, this is a known issue for JWST NIRspec, as the F070LP filter of the G140M grating is affected by second-order contamination at wavelengths longer than $\lambda = 1.27 \, \mu m$. For the G395M grating, they show evidence of a systematic difference in the flux all over the spectral range. To address these issues, we also compare the prism and grating spectra of J-138717 in the lower panels of Fig. \ref{fig:spec}. Differently from the galaxy shown in Fig. 5 of \citet{DR3}, the G140M and prism spectra of J-138717 are consistent with one another along almost all the spectrum, and in particular at $\lambda > 1.27 \, \mu m$, indicating that, for J-138717, the second order contamination is not significant. On the other hand, at $\lambda < 0.9 \, \mu m$ the prism and grating spectra differ significantly, but this difference is likely due to the very low SNR of the G140M spectrum. The flux of the G395M spectrum is systematically offset by $\sim 12\%$, on average, with respect to the prism, which is comparable to the offset reported by \citet{DR3}. This offset is roughly constant along most of the spectrum, while there is some dependence on the wavelength below $\sim 3.4 \, \mu m$ and above $\sim 4.3 \, \mu m$ ($\sim 1.2 \, \mu m$ and $\sim 1.8 \, \mu m$ rest-frame, respectively). 

To further address the flux calibration issues, we considered the independent data reduction of \cite{B24} (updated the 2024 version; hereafter, B24). In Appendix \ref{app:DR3_B24} we discuss in detail the differences between the two spectra (see Fig. \ref{app:DR3_B24_spec} for a direct comparison of the spectra of two reductions). In brief, the G140M spectra of the DR3 and B24 are consistent with each other over the whole wavelength range, although the G140M spectrum of B24 is cut at $1.27 \, \mu m$ ($\sim 4400 \, \AA$ rest-frame), so we can not compare it to the DR3 at longer wavelengths. The G395M spectra of the two reductions are consistent up to $\sim 4 \, \mu m$ ($\sim 1.4 \, \mu m$ rest-frame), while they differ significantly at longer wavelengths. Finally, the prism spectrum shows a clear difference in the slope of the spectrum affecting the full wavelength range. Considering the nature of the problem, it is difficult to determine which is the most reliable data reduction. In this work, we refer to the spectra reduced by the DR3, but we discuss the consistency of the results with those obtained using the spectra reduced by B24 when needed.

\subsection{Templates}\label{sect:templates}

Throughout this paper, we make use of several libraries. For the stellar templates, we considered the MILES \citep{MILES, MILES11} and IRTF + IRTF-extended \citep{IRTF0, IRTF} libraries. The MILES library consists of 985 stars observed in the wavelength range $\sim 3500 - 7500 \, \AA$, with an average resolution R$\sim 2000$. The IRTF and IRTF-extended libraries consist of a total of 385 stars with different wavelength ranges. For this paper, we use a version of the library where all stars have been rebinned to a common wavelength range $\sim 0.8 - 5.2 \, \mu m$, having an average resolution of R $\sim 2000$ from 0.8 to 2.5 $\mu m$ and R $\sim 2500$ from 2.5 to 5.2 $\mu m$ (see \citealt{Rock+17} and \citealt{FLB+24} for details); in the following, we refer to this version of the library simply as IRTF. As for the libraries of simple stellar population (SSP) models, we considered the following.
\begin{enumerate}[i)]

\item The EMILES models \citep{EMILES}. These models cover the wavelength range $1680 - 50000 \, \AA$ and are based on fully empirical stellar libraries, namely the NGSL library \citep{NGSL} in the ultraviolet, the MILES \citep{MILES, MILES11}, the Indo-US \citep{IndoUS}, and CaT \citep{CaT} libraries in the optical/NIR, and the IRTF library \citep{IRTF} in the NIR/IR, with a spectral resolution of 2.51 $\AA$ full-width at half maximum (FWHM) in the optical and $60$ km s$^{-1}$ in the NIR (see \citealt{EMILES} for details). For the present work, we make use of an updated version of the models that differs from that described in \citet{EMILES} only in the spectral range $\sim 0.82 - 2.5 \mu m$, where 205 additional stellar spectra from the extended IRTF library have been included, on top of the previous ones, while the computation of the models follows the same prescriptions and ingredients as for the original EMILES models (Vazdekis et al., in prep.). They have been recently used in \citet{FLB+24}. 

\item The X-Shooter Library of SSP models (\citealt{XSL}, hereafter XSL) is entirely based on the X-Shooter stellar library \citep{XSL_stars}, covering the wavelength range $0.3 - 2.5 \, \mu m$, with a spectral resolution R $\sim 10000$.

\item The \citet{CvD18} models (hereafter CvD18) models are the updated version of the \citet{CvD12} models. They cover the wavelength range $0.35 - 2.5 \, \mu m$ and are based on the MILES and extended IRTF stellar library \citep{IRTF} in the optical and NIR, respectively, having a fixed instrumental resolution of $\sigma_{\rm inst} = 100$ km s$^{-1}$. One key feature of these models is that they include models with non-solar values in the abundance ratios for different elements. 

\item The \citet{M05} models (hereafter M05) are semi-empirical models having a prescription for evolved stars determined by the fuel consumption theorem of \citet{RV81}. They are based on the BaSel library \citep{basel} and the \citet{Lancon00} empirical library of TP-AGB stars, covering a wide range of wavelengths, $91 - 1.6 \times 10^6 \, \AA$, and having a relatively low resolution, R$\sim 100$. In the rest-frame wavelength range covered by J-138717, $0.35 - 1.4 \, \mu m$, this corresponds to a FWHM varying from $10 \, \AA$ to $50 \, \AA$. 

\item The Maraston (2013, described in \citealt{M13}; hereafter, M13) models are an updated version of the M05 models, adopting the same libraries and having the same instrumental characteristics. The only two differences are: (i) they assume a different fuel consumption, resulting in a heavy contribution of TP-AGB stars for the M05 models and a mild contribution for the M13 models; (ii) the isochrones used to compute the M13 models have been re-calibrated in the age range $0.2-1$ Gyr by fitting the Color-Magnitude diagram of the Magellanic Cloud star clusters.

\end{enumerate}

\begin{table}
	\centering
	\caption{Relevant properties of the SSP models libraries}
	\label{tab:libraries}
	\begin{tabular}{|c|c|c|c|c|}
		\hline
		& & & Age & [M/H] \\
		Library & Isoch. & IMF & range & range   \\
		 &  & & (Gyr)  & (dex)  \\
		\hline
		& & & &\\
		EMILES & BaSTI & Bi1.3 & $(0.03, 3.5)$ & $(-1.79, 0.26)$\\
		CvD18 & MIST & Kroupa & $(1.0, 3.0)$ & $(-1.00, 0.20)$\\		
		XSL & PARSEC & Kroupa & $(0.05, 3.2)$ & $(-1.80, 0.20)$\\
		M05 & Padova & Kroupa & $(10^{-6}, 3.0)$ & ($-1.35, 0.35)$\\
		M13 & Padova & Kroupa & $(10^{-6}, 3.0)$ & ($-1.70, 0.35)$\\
		& & & &\\
		\hline
	\end{tabular}
	\begin{minipage}{\columnwidth}
	Columns: (1) name of the SSP models library. (2) Isochrones used in the models considered in this work. (3) IMF of the models adopted in this work. (4) Age ranges of the model libraries. We note that there are no templates with ages below 1 Gyr available in the CvD18 library. (5) Metallicity ranges of the models libraries. 
	\end{minipage}

\end{table}

We note that all these models are computed for different isochrones and IMF types, and are available for large ranges of ages and metallicities. Here, we considered the BaSTI isochrones \citep{basti} for EMILES, the PARSEC isochrones \citep{PC0, PC1, PC2} for XSL, the MIST isochrones \citep{MIST1, MIST2} for CvD18, and the Padova isochrones \citep{Padova00} for M05 and M13. As for the IMF, we adopted the \citet{Kroupa2001} IMF for all the libraries, except for the EMILES models for which we used a bimodal IMF with logarithmic slope $\Gamma_{\rm b} = 1.3$, which resembles the Kroupa IMF. As for the maximum age of the models we considered the age of the models closest to the age of the Universe at the redshift of the J-138717 ($\sim 3.4$ Gyr). We also constrain the metallicity between [M/H]$= -1.80$ and $0.35$ dex. We verified that including models with older ages and lower and higher metallicities has a negligible impact on the results, while increasing the computational time. In table \ref{tab:libraries} we summarize the relevant stellar population properties of the libraries adopted in this work.

\section{Physical and stellar population parameters estimate}\label{sect:fits}

\subsection{Integrated stellar velocity dispersion}\label{sect:kinematics}

We estimated the stellar velocity and velocity dispersion ($\sigma_*$) by fitting the grating spectra G140M and G395M, separately, using \texttt{pPXF} (\citealt{ppxf_2004, ppxf_2017, ppxf_2023}, updated to version 9). To account for residuals in the background subtraction during the data reduction, we included additive polynomials in the fit. To choose the degree of the polynomials, we initially performed a fit of the spectrum using additive polynomials of degree 10. From this fit, we computed the standard deviation of the residuals ($\sigma_{\rm std}$) and masked all the spectral pixels deviating more than $3\sigma_{\rm std}$ from the best fit. Then, we iteratively fitted the spectrum with additive polynomials at increasingly higher degrees, starting from 2. For each iteration (degree), we computed the best-fit velocity and velocity dispersion and estimated the relative errors (see below). As a result, we found that the best-fitting velocity and velocity dispersions are consistent within the errors for the order of the polynomials larger than 10 and 12 for the G140M and G395M spectra, respectively. For this reason, as our fiducial estimate for the velocity dispersion and errors of J-138717 we considered the best-fitting solutions found using additive polynomials of degree 10 and 12 for G140M and G395M, respectively. To assess the errors, for each iteration we performed a wild bootstrapping of the residuals by resampling the best-fit spectrum with the noise Gaussianly distributed from the residuals. For each iteration, we performed 100 realizations and considered the standard deviation as the relative error. 

As templates to run \texttt{pPXF}, we used both the stellar libraries and SSP models. We used the MILES stars to fit the G140M spectrum, corresponding to the optical rest-frame, and the IRTF library to fit the G395M spectrum, corresponding to the NIR rest-frame spectrum, because the MILES (IRTF) stars do not cover the NIR (optical) rest-frame wavelength range. As SSP models, we considered the EMILES, XSL, and CvD18 models. We did not consider the M05 and M13 models because they have lower resolutions than the gratings. We note that the IRTF, XSL, and CvD18 libraries are affected by tellurics at $\sim 1.3 - 1.4 \, \mu m$ \citep{XSL, IRTF}, so we masked this rest-frame wavelength region when performing the fit for the G395M spectrum with these templates.

All estimates are consistent within the errors with average velocity dispersion $\sigma_* = 198$ km s$^{-1}$ and standard deviation of 10 km s$^{-1}$, which we adopt as our reference value and associated uncertainty. The systemic velocities are also consistent and correspond to a spectroscopic redshift $z_{\rm spec} \simeq 1.8845$ (lower than $z_{\rm phot} = 2.16$) that we manually verified by checking the wavelengths of the most prominent emission and absorption lines.

\subsection{Stellar population properties}\label{sect:agemet}

\begin{figure}
\includegraphics[width=\columnwidth]{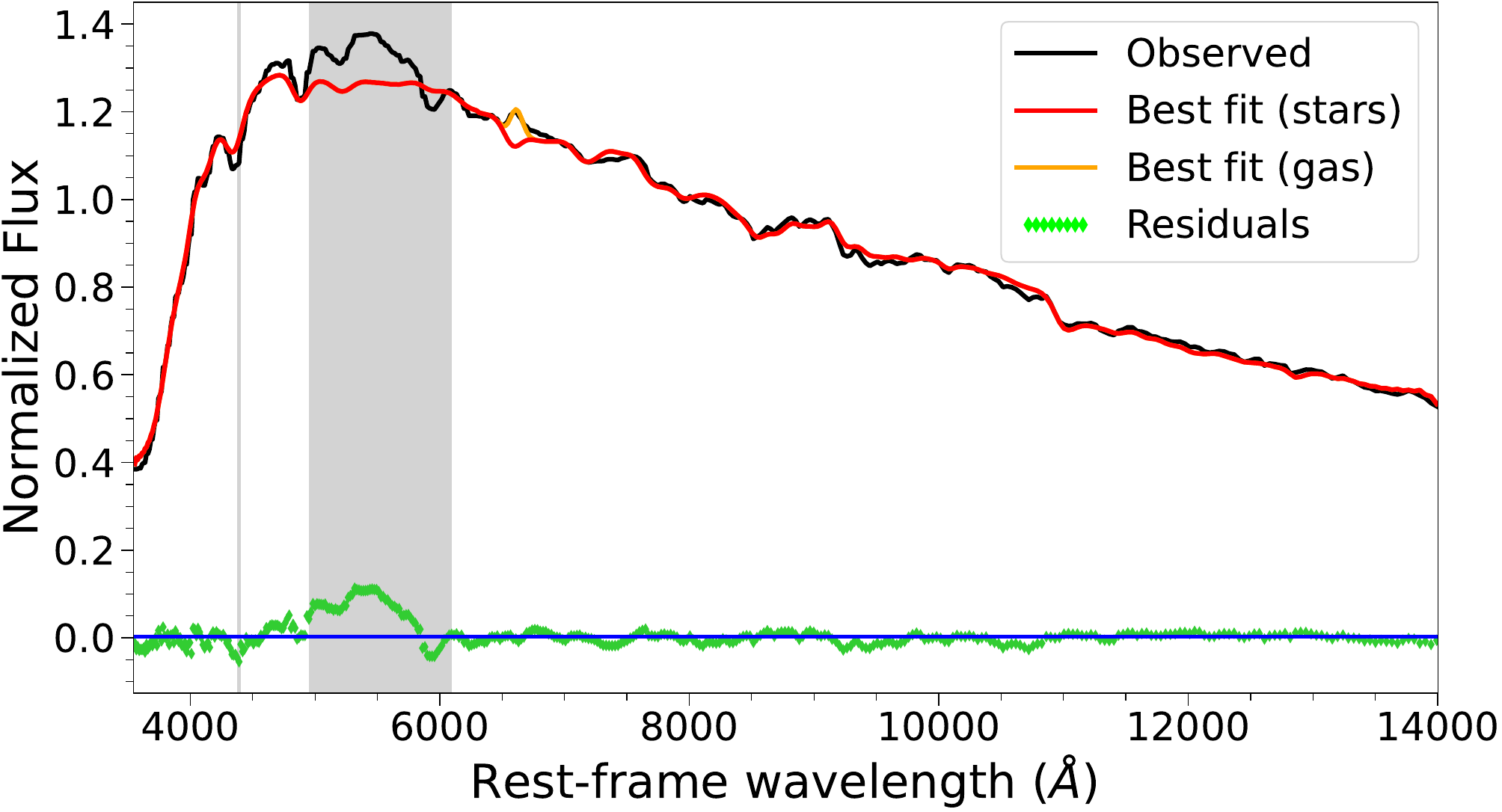}
\includegraphics[width=\columnwidth]{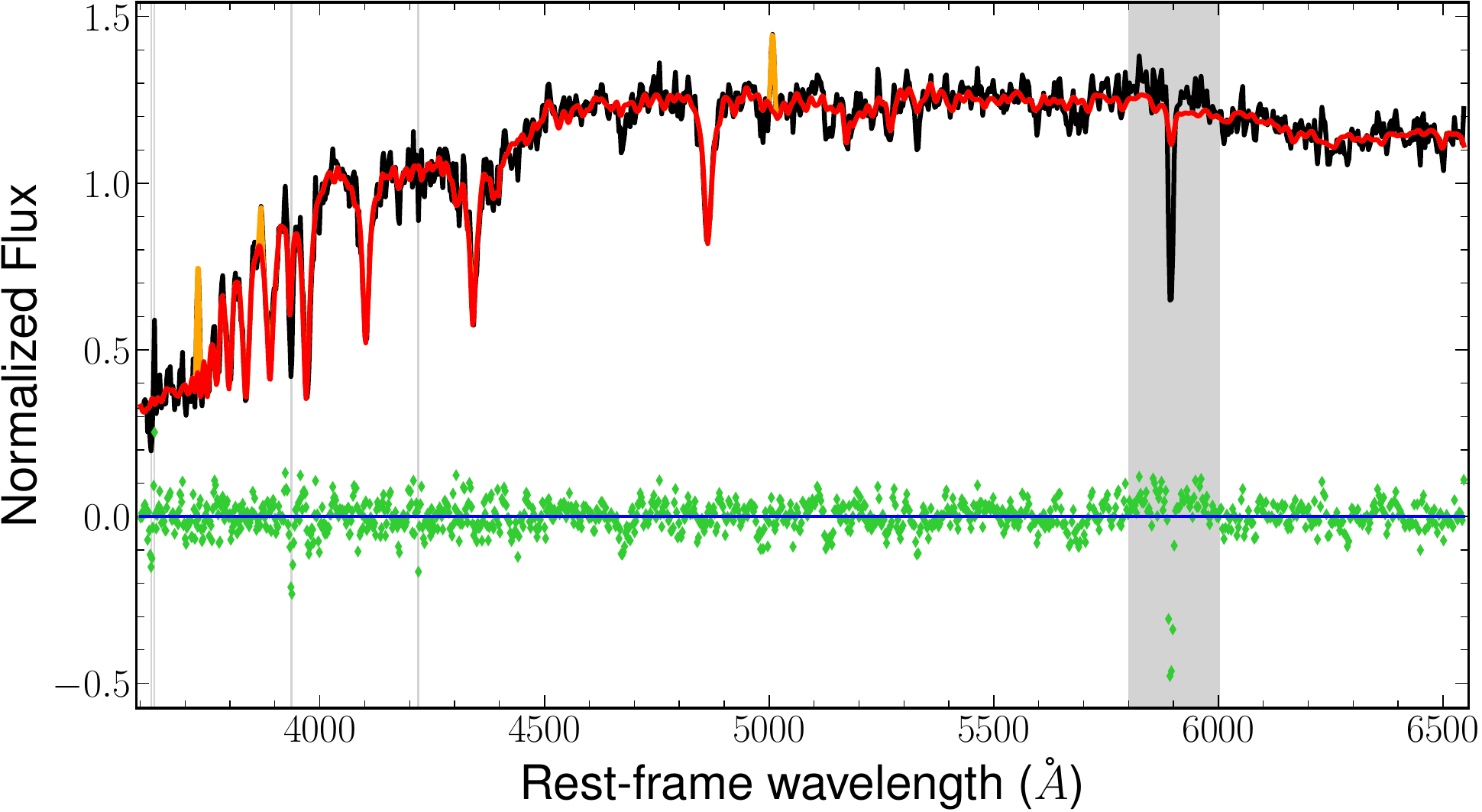}
\includegraphics[width=\columnwidth]{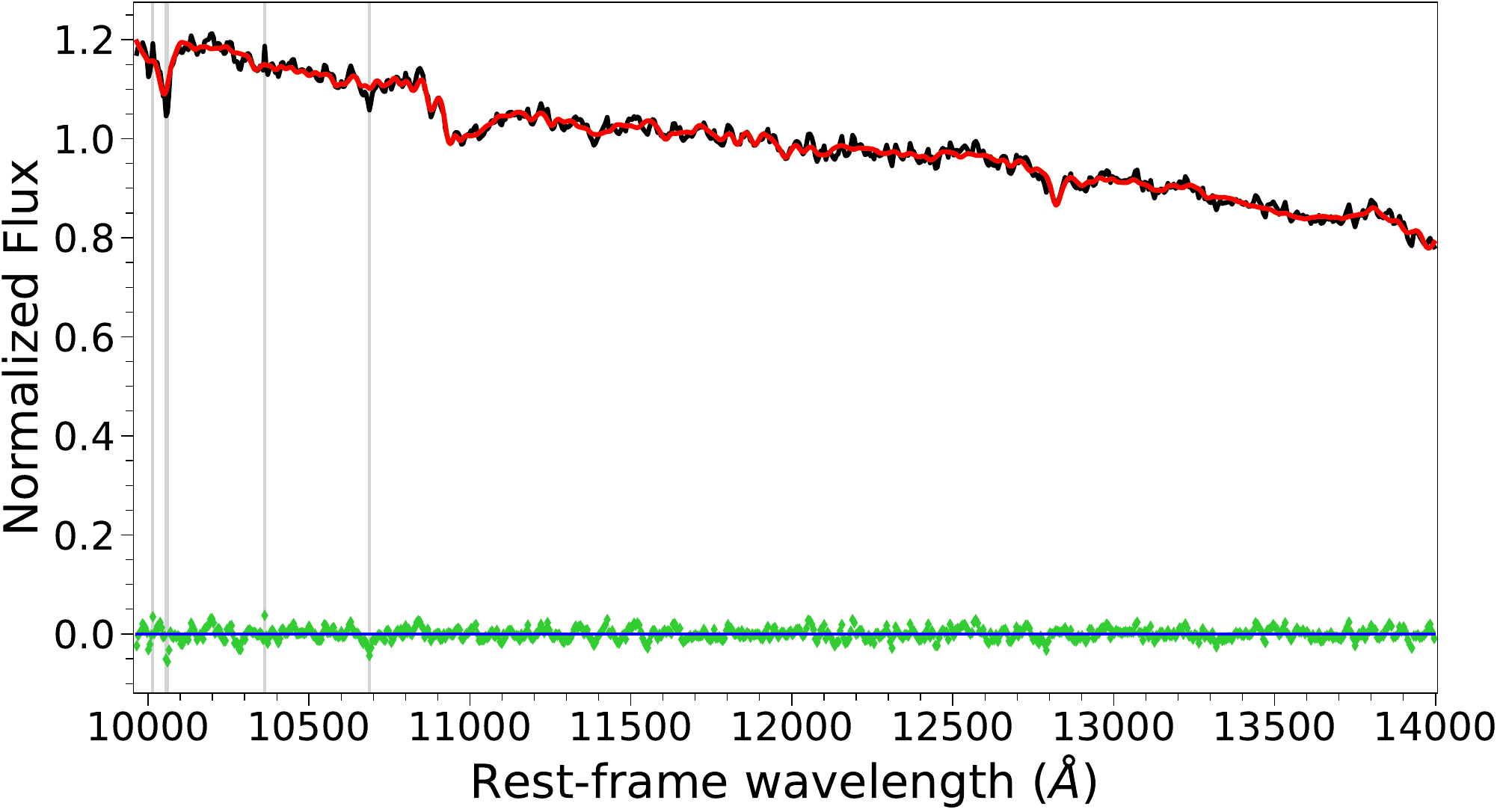}
\caption{Best-fit of the prism (upper panel), G140M (middle panel), and G395M (lower panel) of the EMILES models. The three spectra are fitted separately. In each panel, the black line is the rest-frame observed spectrum. The red line is the best-fit to the stellar continuum. The orange line is the best-fit to the emission lines. The grey shaded regions are excluded from the fit. The green diamonds are the residuals and the blue horizontal line is their median value. }
\label{fig:bestfit}
\end{figure}

\begin{figure}
\includegraphics[width=\columnwidth]{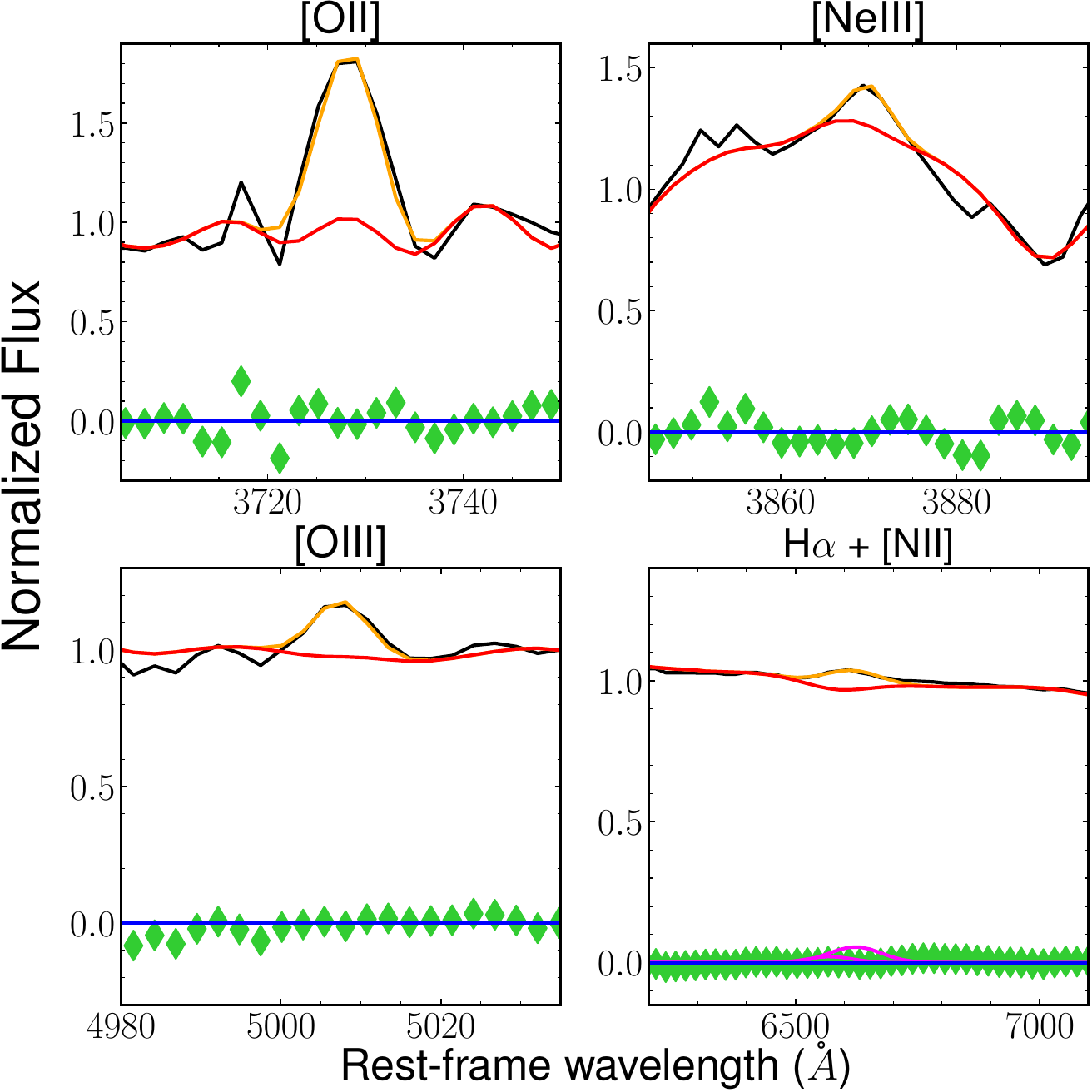}
\caption{Zoom-in of the best-fits for the emission lines. The lines and the symbols are the same as in Fig. \ref{fig:bestfit}. The magenta line in the lower right panel show the best-fit templates for the H$\alpha$ and [NII] lines separately.}
\label{fig:emlines}
\end{figure}

\begin{figure*}
\centering
\includegraphics[width=0.24\textwidth]{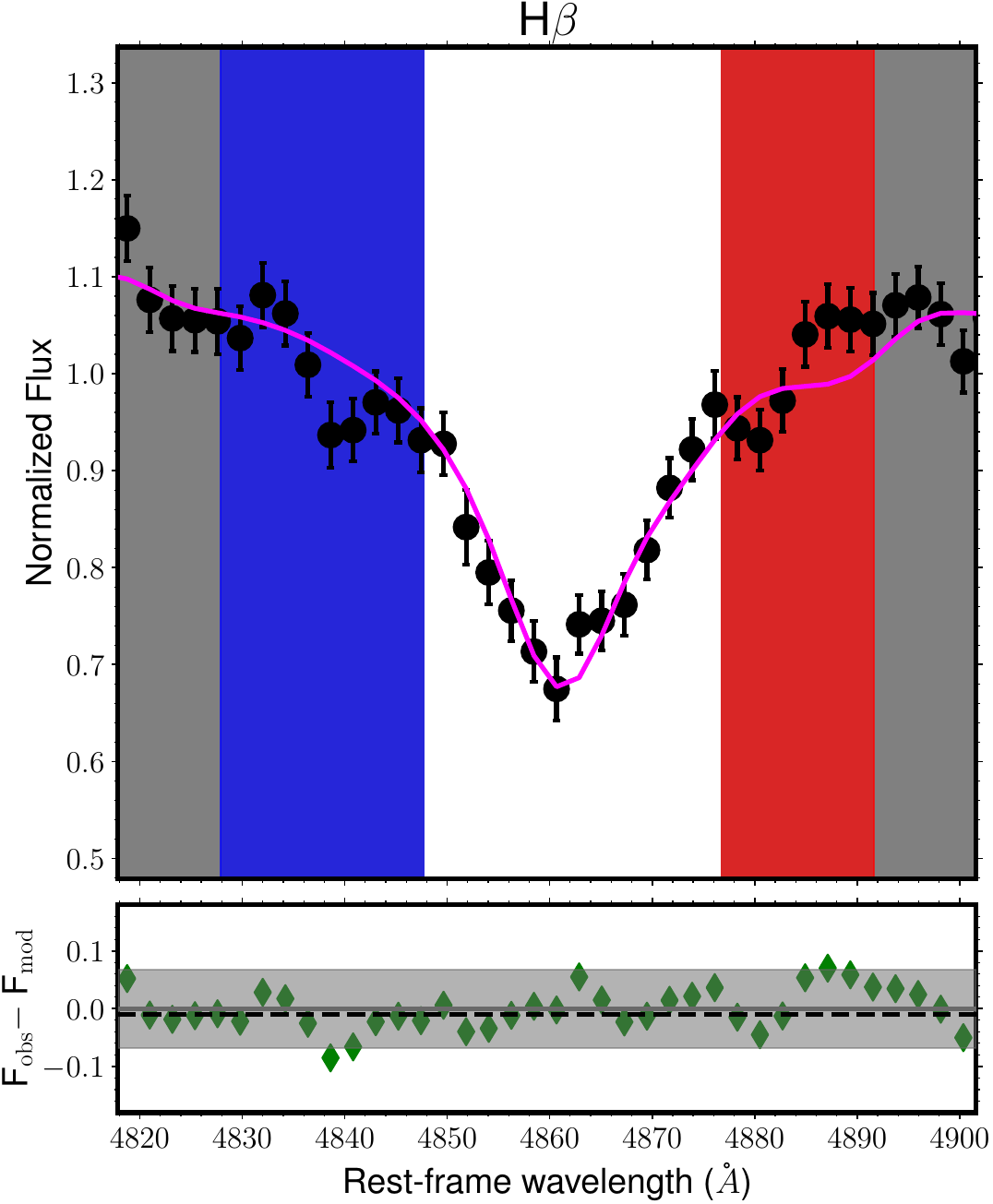}
\includegraphics[width=0.24\textwidth]{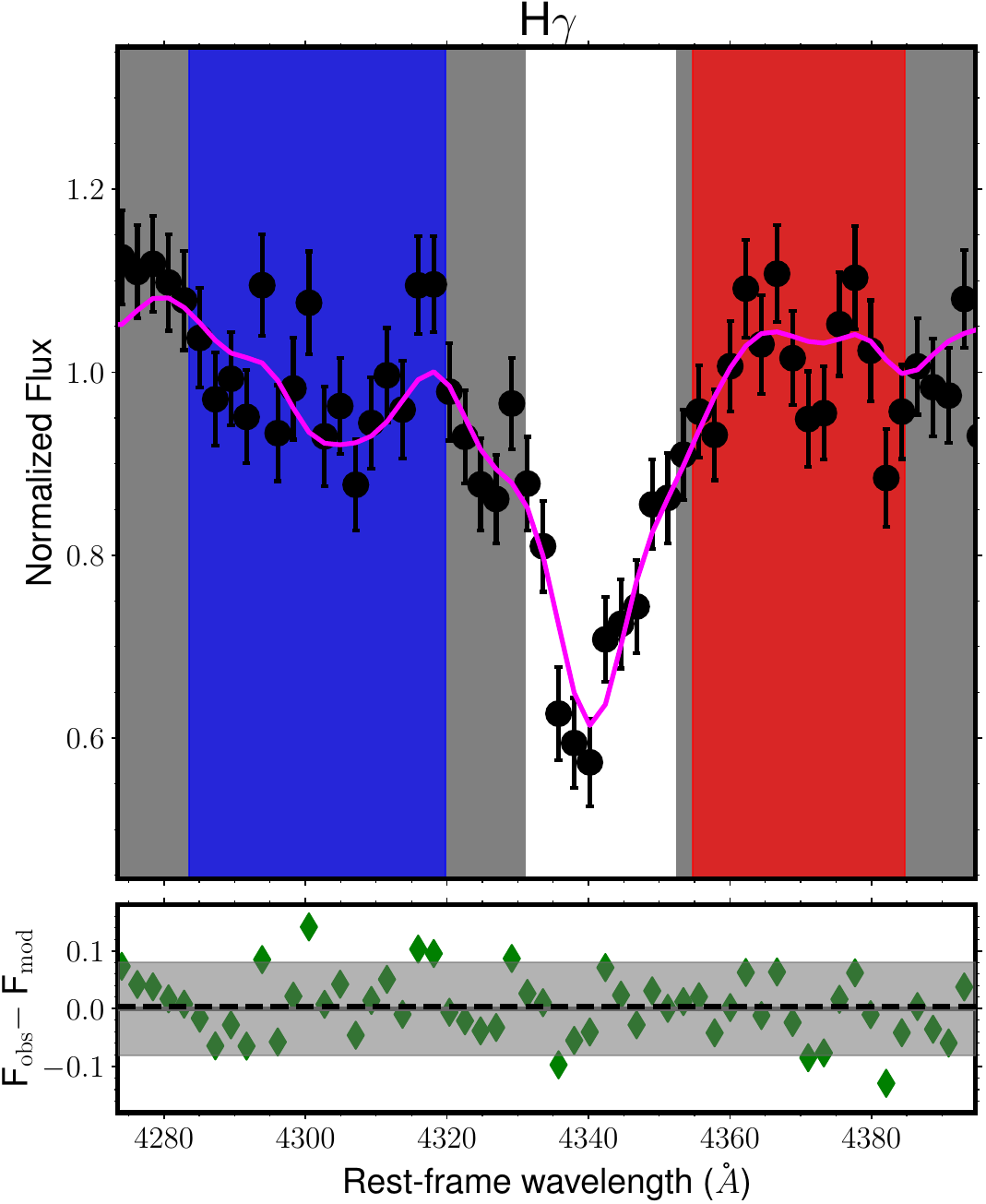}
\includegraphics[width=0.24\textwidth]{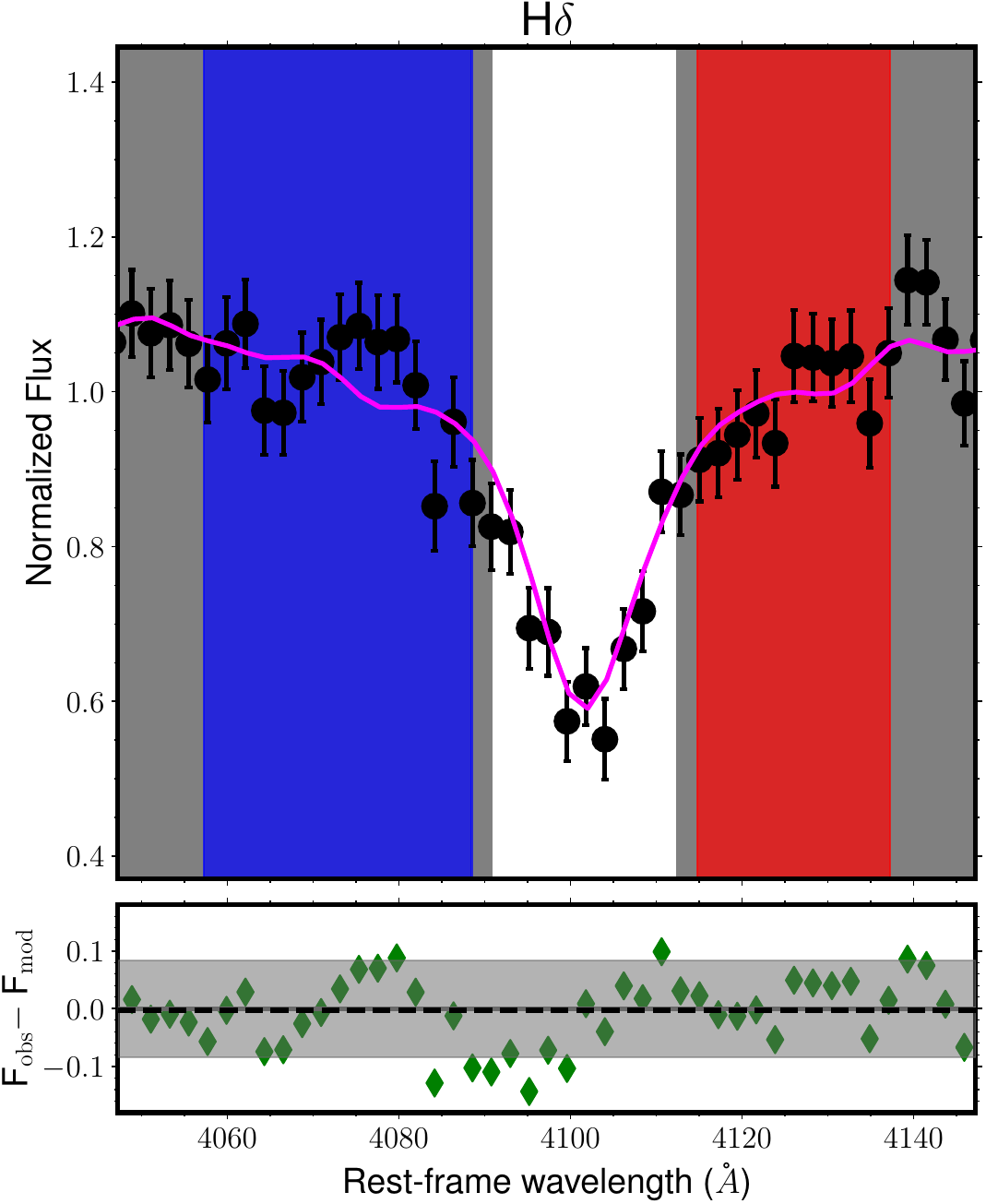}
\includegraphics[width=0.24\textwidth]{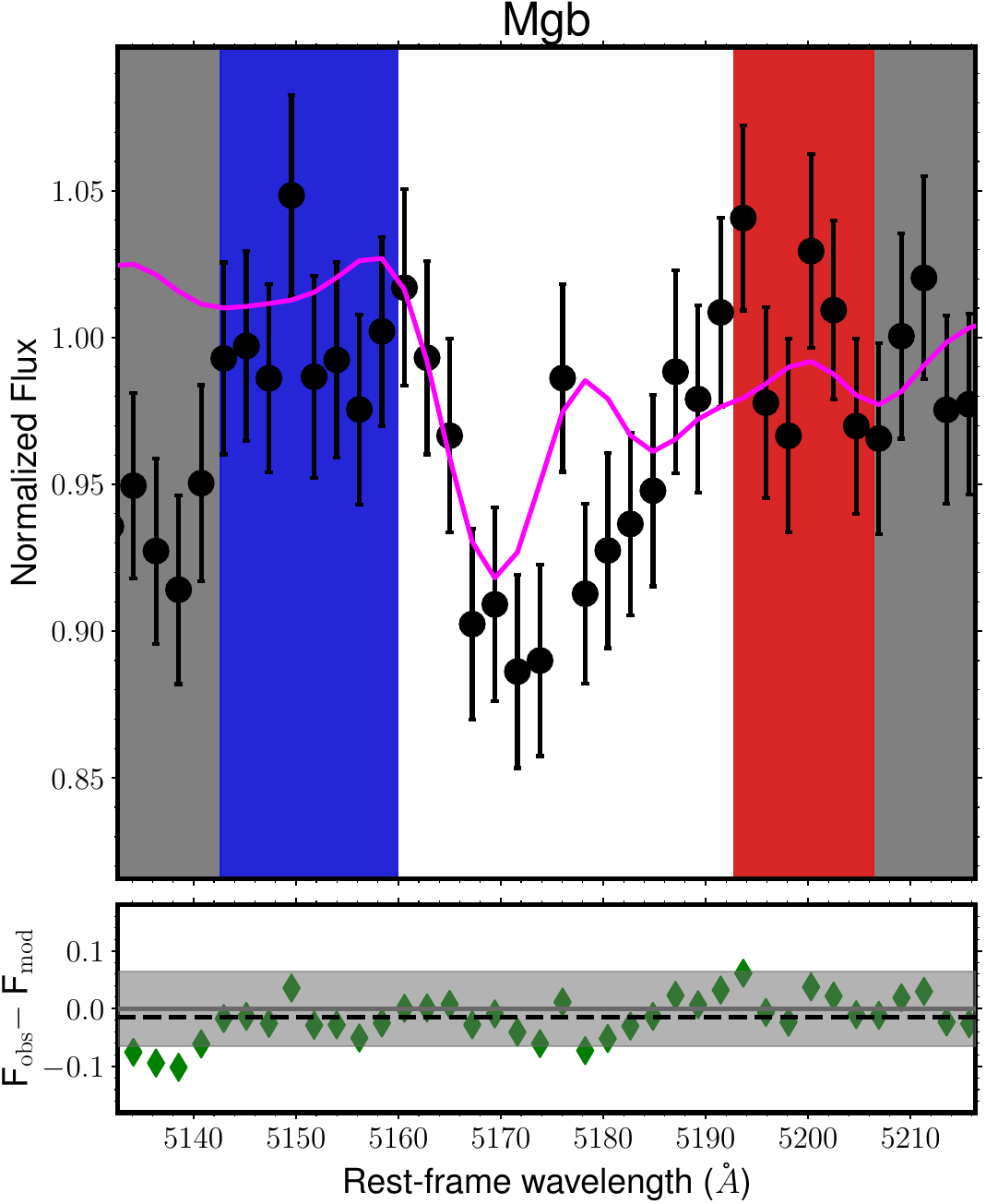}
\includegraphics[width=0.24\textwidth]{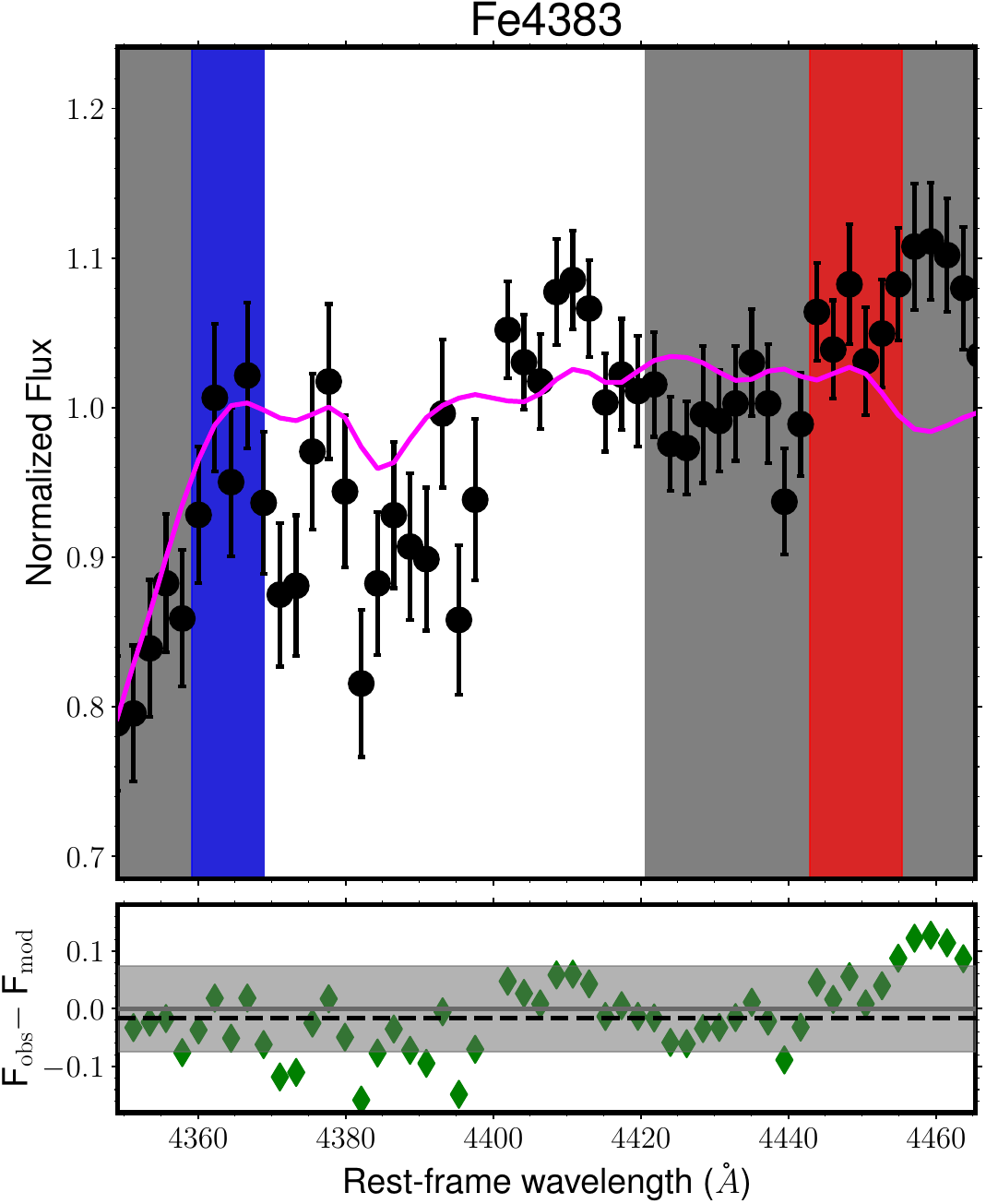}
\includegraphics[width=0.24\textwidth]{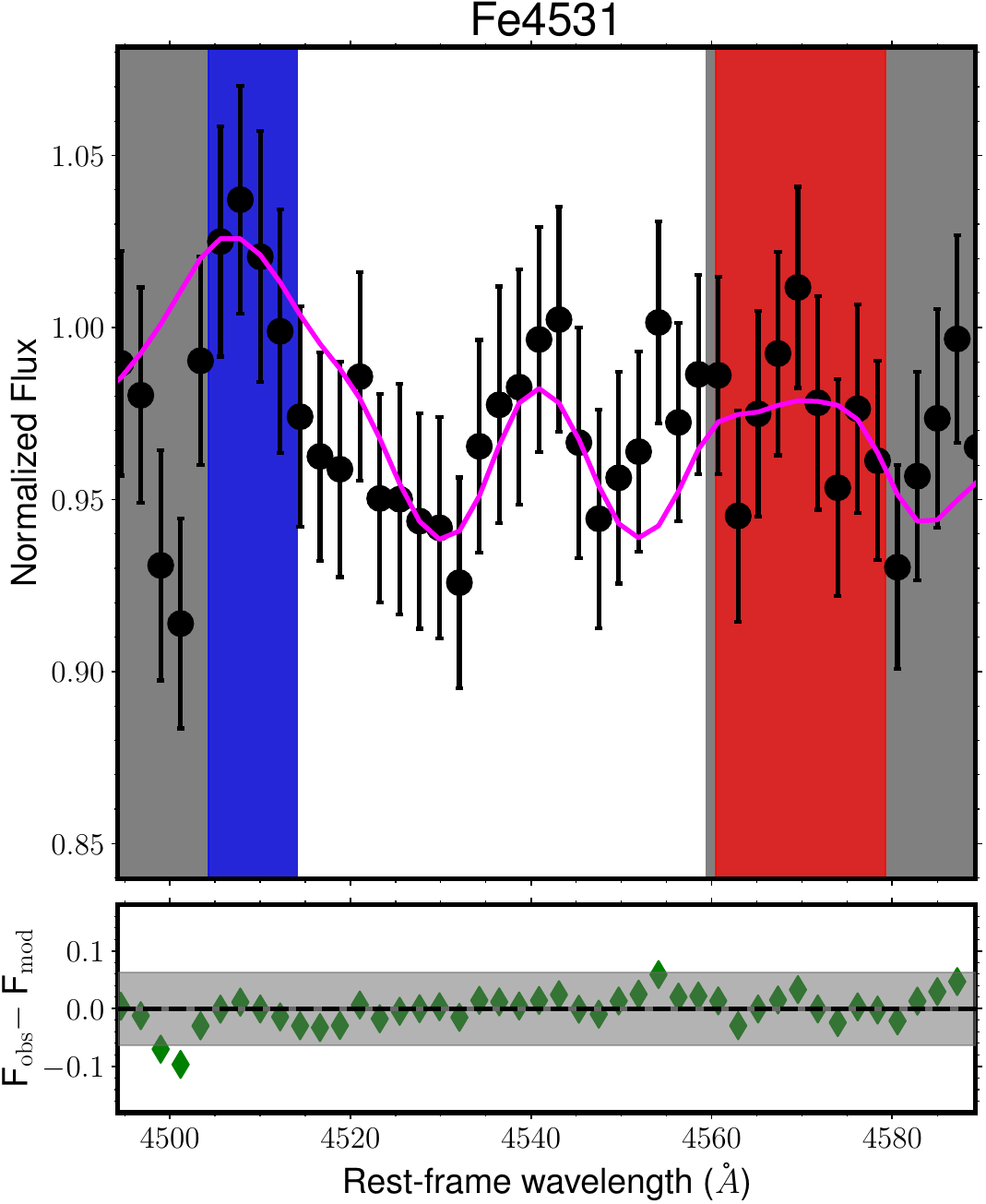}
\includegraphics[width=0.24\textwidth]{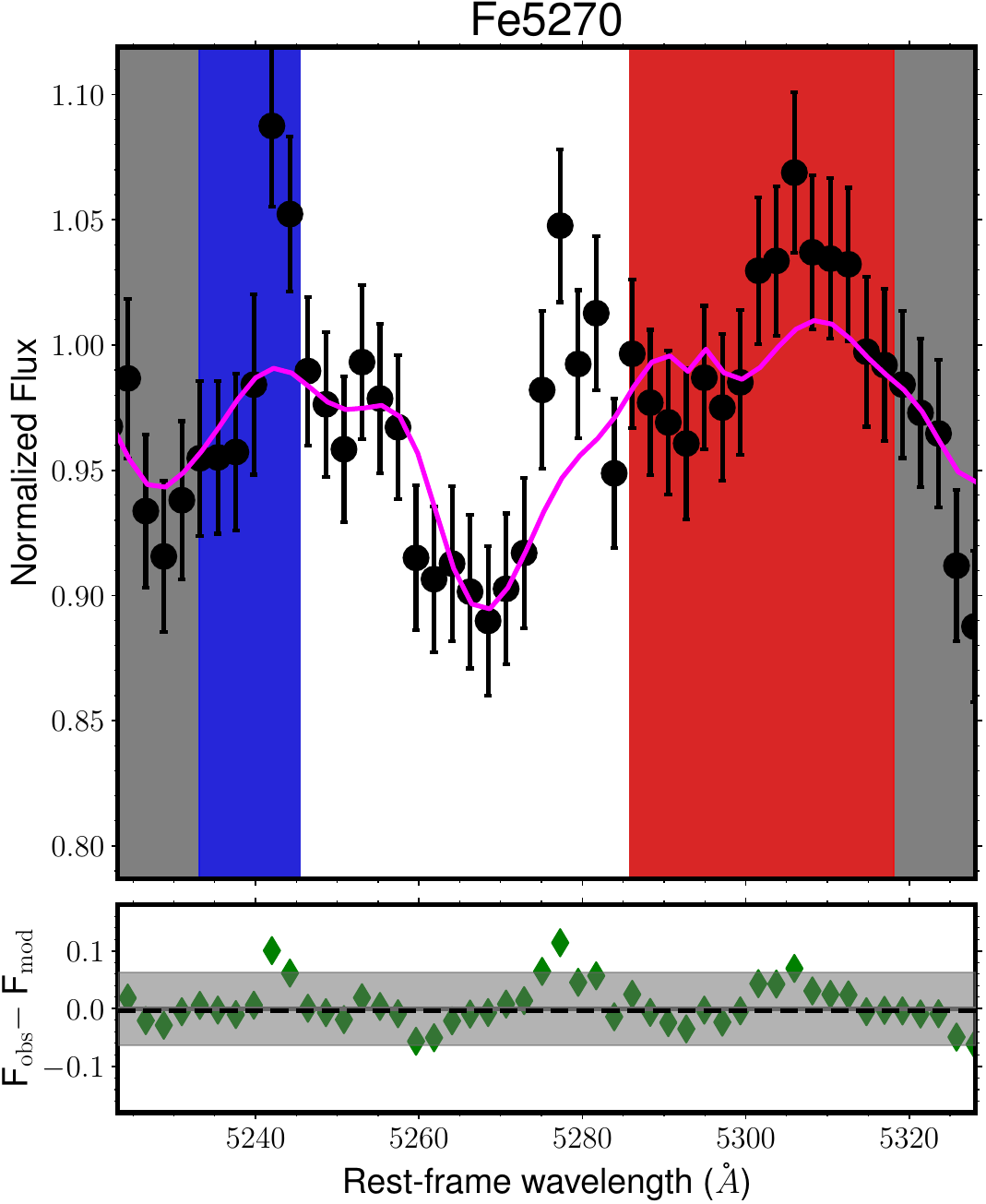}
\includegraphics[width=0.24\textwidth]{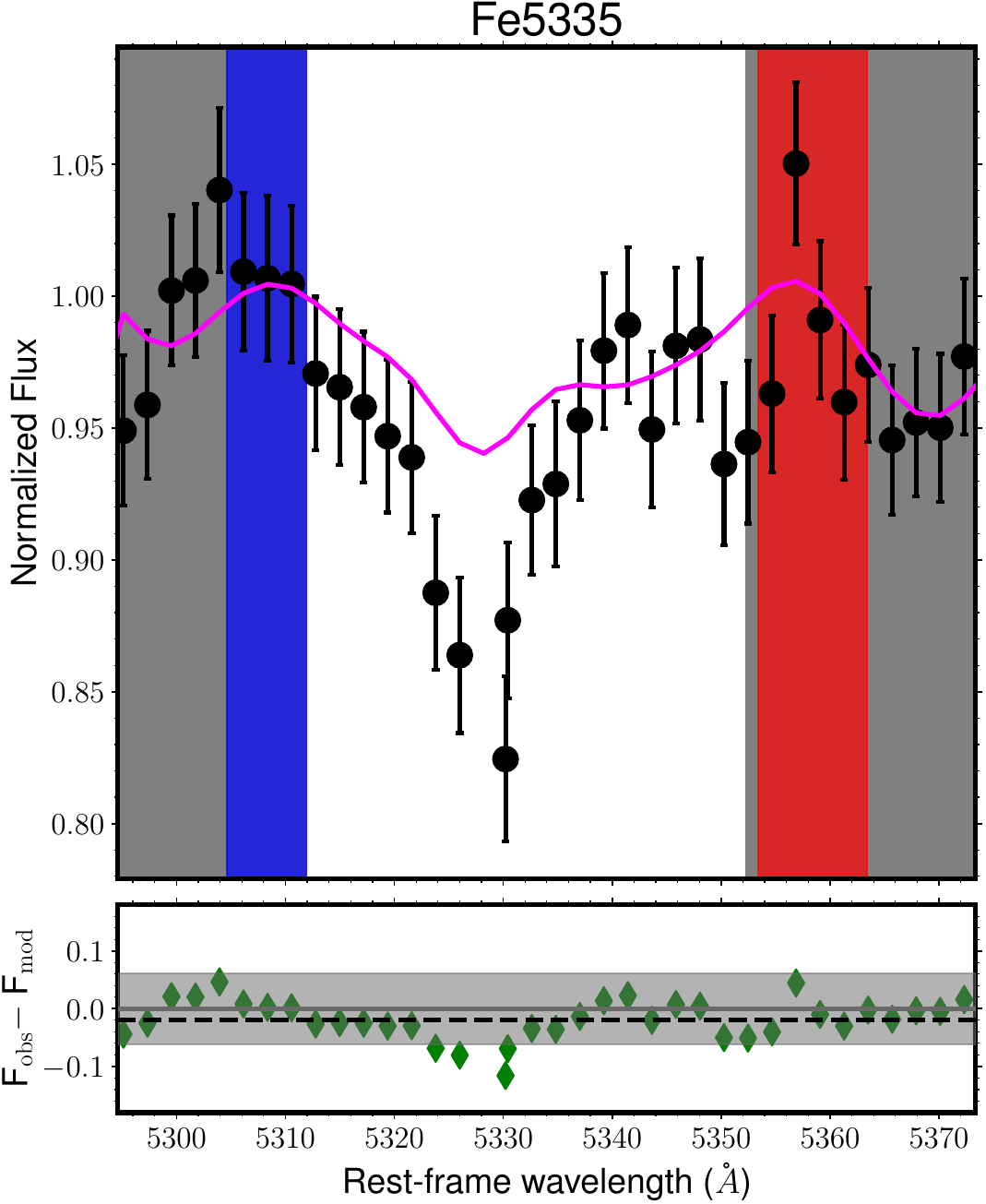}
\caption{Comparison of the observed indices (black dots are the spectral pixels) with the predictions of the FIF best-fit models (magenta line), having an age of 0.81 Gyr and a metallicity of $-0.19$ dex. The blue and red regions correspond to the wavelengths where the pseudo-continua are measured and the white regions are the wavelengths where the index is measured. In the bottom panels the residuals are plotted as green diamonds. The grey lines and shaded regions are the median of the residuals and the average observed errors on the indices, respectively, while the black dashed line marks the 0. }
\label{fig:fif}
\end{figure*}

To estimate the stellar age and metallicity of J-138717, we performed full-spectral fitting using \texttt{pPXF} on the G140M, G395M, and prism spectra separately. For the G140M spectrum, we fitted the rest-frame wavelength range $\lambda \sim 3600 - 6500 \, \AA$, masking out the region around the NaD line ($\lambda\sim 5900$) that is not well reproduced by stellar population models, as its strength is determined by neutral gas absorption; we further discuss NaD in Sect. \ref{sect:c2nad}. We excluded from the fit the NUV region because of the low SNR ($\lesssim 5 \, /\AA$, on average). For the G395M spectrum, we fitted the rest-frame wavelength range $\lambda\sim 1 - 1.4 \, \mu m$, since at longer wavelengths the flux calibration is uncertain, as discussed in section \ref{sect:data}. For consistency with the two gratings, we restricted the fit of the prism to the rest-frame spectral region $\sim 0.36 - 1.4 \, \mu m$. For the prism, we had to mask the region $5000-6000 \, \AA$ (rest-frame), because it can not be fitted simultaneously with the other wavelength ranges by any model library (or even other fitting codes). We note that we can fit the $5000-6000 \, \AA$ region when considering the G140M spectrum. Since the G140M spectrum is consistent with that of the prism in that wavelength region, while there is a systematic offset with respect to the G395M spectrum (see Fig. \ref{fig:spec}), we attribute the mismatch to flux calibration issues of the observed spectrum in the NIR. We further discuss this problem in section \ref{sect:caveats}.

For the G140M spectrum, we fitted a two-parameters (A$_V$, $\delta$) attenuation law for the dust, as modeled in \citet{ppxf_2023} (eq. 23) and multiplicative polynomials of degree 8. For the G395M spectrum, we only fitted multiplicative polynomials of degree 14; here, we do not include a dust attenuation law in the fit as the NIR rest-frame spectrum is not significantly affected by the dust, and we verified that including it does not change significantly our results. Analogously to the additive polynomials described in section \ref{sect:kinematics}, we chose the orders of the multiplicative polynomials by iteratively increasing the degree during the fit until it converged to a constant, consistent solution. For the prism, we fit the dust attenuation but we do not include multiplicative polynomials due to the relatively low number of spectral pixels. We verified that including them does not affect significantly the results. For the G140M and prism spectra, we simultaneously fitted the stars and the gas emission lines of the [OII] doublet ($\lambda = 3726, 3728 $ \AA), [NeIII] ($\lambda = 3869$ \AA) [OIII] ($\lambda = 5007$ \AA), and the Balmer lines. For the prism spectrum, we also fitted the [NII] line ($\lambda = 6584$ \AA) that is blended with H$\alpha$, as we discuss in the next sections. When fitting the G140M and G395M spectra we constrained the stellar velocity dispersion to be within 3 standard deviations from our reference value (i.e., within $\sigma_* = 198\pm30$ km s$^{-1}$; Sect. \ref{sect:kinematics}). On the other hand, we let \texttt{pPXF} fit the kinematics of the prism, since the instrumental resolution is much lower than $\sigma_*$.

We then estimated the best-fitting models as follows. For each spectrum, we performed a first fit and computed the standard deviation of the residuals ($\sigma_{\rm std}$). Then, we performed a second fit and masked all the spectral pixels deviating $>3\sigma_{\rm std}$. We consider the results of this second fit as our best fit. The best fits of the three spectra are shown in Fig. \ref{fig:bestfit}. 

We calculate the light-weighted ages and metallicities from the best-fits as:

\begin{equation}\label{eq:age}
{\rm log}_{10}{\rm Age} = \frac{\sum_i  w_i \rm{log_{10} Age}_i}{\sum_i w_i}
\end{equation}

\begin{equation}\label{eq:met}
{\rm [M/H]} = \frac{\sum_i w_i {\rm [M/H]}_i}{\sum_i  w_i}
\end{equation}

\noindent where $w_i$ is the weight of the $i$-th input template and the sums are performed over all the templates. To estimate the errors of the age, metallicity, and dust attenuation parameters, we performed a wild bootstrapping of the residuals. Specifically, we performed 100 realizations of each spectrum by shuffling the residuals over windows $\sim300 \, \AA$ wide, to account for wavelength correlated errors, and then re-distributing (i.e., adding) the shuffled residuals to the best-fitting spectrum Gaussianly. For each realization, we estimated the age and metallicity and took the standard deviations of all the realizations as the errors. Using the EMILES models, from the fits of the optical (G140M), NIR (G395M), and prism spectrum we estimate an age of $0.91\pm0.21$, $0.95\pm0.25$, and $0.89\pm0.11$ Gyr, respectively, and a metallicity of $-0.28\pm0.17$, $-0.38\pm0.15$, and $-0.35\pm0.08$ dex. The attenuation parameters estimated from the optical and prism spectrum are A$_V = 0.65\pm0.18, 0.63\pm0.05$ mag and $\delta = -0.82\pm0.29, -0.90\pm0.12$, respectively. In Fig. \ref{fig:emlines} we show the best fits to the emission lines. We estimate $F(\mbox{H}\alpha) = 4.7\pm0.5\times 10^{-19} \mbox{erg cm}^{2} \, \mbox{s}^{-1} $, $F(\mbox{[NII]}) = 1.5\pm0.1\times 10^{-18} \mbox{erg cm}^{2} \, \mbox{s}^{-1} $, $F([\mbox{OII}]) = 5.2\pm0.5\times 10^{-19} \mbox{erg cm}^{2} \, \mbox{s}^{-1} $, $F([\mbox{NeIII}]) = 1.7\pm0.5\times 10^{-19} \mbox{erg cm}^{2} \, \mbox{s}^{-1} $, and $F([\mbox{OIII}]) = 4.0\pm0.2\times 10^{-19} \mbox{erg cm}^{2} \, \mbox{s}^{-1} $; except H$\alpha$, the Balmer lines are consistent with having zero flux in emission. 

In Appendix \ref{app:DR3_B24_agemet}, we show the results of the fits with the other SSP libraries. Overall, the stellar population parameters are consistent when using different libraries or fitting different spectral ranges. In Appendix \ref{app:DR3_B24_agemet}, we also show how different the stellar population properties are when fitting the B24 prism spectrum. In general, the latter provides older ages and higher metallicities, while the attenuation parameters are consistent. This difference is expected since the B24 prism spectrum is flatter than that of the DR3. On the other hand, we find fully consistent results when fitting the G140M and G395M spectra for all SSP libraries.

Since the stellar continuum is attenuated by dust, possibly affecting the age and metallicity estimates, we re-derive these parameters using the Bayesian full-index fitting (FIF) method \citep{FIF1, FIF2}. The FIF method is a hybrid approach between the full-spectral fitting and the line-strength analysis. The basic idea is to use as much spectral information as possible, like in the full-spectral fitting, but to limit the fit to a number of carefully selected spectral indices, based on their sensitivity to the stellar population parameters to be determined. For our science case, the key advantage of the FIF is that it relies on the spectral indices, thus minimizing the source of uncertainties affecting the stellar continuum such as the dust absorption or flux calibration issues. 

The FIF method performs a pixel-by-pixel fit of all the spectral pixels within the band-passes of the spectral features fitted. In this work, we use the \texttt{emcee} Bayesian Monte-Carlo-Markov-Chain sampler~\citep{emcee} to maximize the following likelihood function:
\begin{equation}
\ln(\textbf{O}|\textbf{S}) = - \frac{1}{2}\sum_{n}\left[\frac{(O_n-M_n)^2}{\sigma_n^2} - \ln\left( \frac{1}{\sigma^2_n}\right) \right] \; ,
\end{equation}
where the sum extends over all the spectral pixels within the band-passes fitted, with $O_n$ and $M_n$ being the observed and the model fluxes, respectively, of the $n$-th spectral pixel, while $\sigma_n$ is the uncertainty on $O_n$. As for the spectral indices, we considered H$\beta$, H$\gamma$, and H$\delta$, which are primarily sensitive to the galaxy's age, and the Fe4383, Fe4531, Fe5270, and Fe5335, which are sensitive to the metallicity. We also include the Mgb, which is sensitive to both metallicity and [$\alpha$/Fe] abundance. To estimate the stellar population properties we adopted the EMILES models, and we verified that the XSL library provides consistent results. Given that the FIF is based on lower statistics with respect to the full-spectral fitting and does not rely on the stellar continuum, we do not fit the M05 and M13 models, as they require to reduce the resolution of the observed spectrum, and neither the CvD18 models, as they are not computed at ages lower than 1 Gyr that would thus affect the metallicity.

The FIF best-fitting age is $0.81^{+0.05}_{-0.07}$ Gyr, consistent with the results of the full-spectral fitting, with a metallicity of $-0.19^{+0.11}_{-0.07}$ dex, which is consistent with the estimate of the full-spectral fitting of the optical spectrum, albeit $\sim 0.1$ dex higher ($\sim 0.2$ with respect to the fit of the NIR and prism). In Fig. \ref{fig:fif} we show how the observed indices compare to the best-fit model (i.e. with the estimated median values of age and metallicity). The Balmer lines, which are more sensitive to the age, are well-fitted. The iron lines Fe4531 and Fe5270 are also well reproduced, while Fe4383, Fe5335, and Mgb are underestimated. We verified that when including other spectral indices sensitive to the metallicity (CN1, CN2, Ca4227, CN3883) the metallicity is slightly lower ($-0.23^{+0.07}_{-0.11}$), yet still systematically higher than the one calculated from the full-spectral fitting. 

\subsection{Mass, SFR, and SFH}\label{sect:mass}

\begin{figure}
\includegraphics[width=\columnwidth]{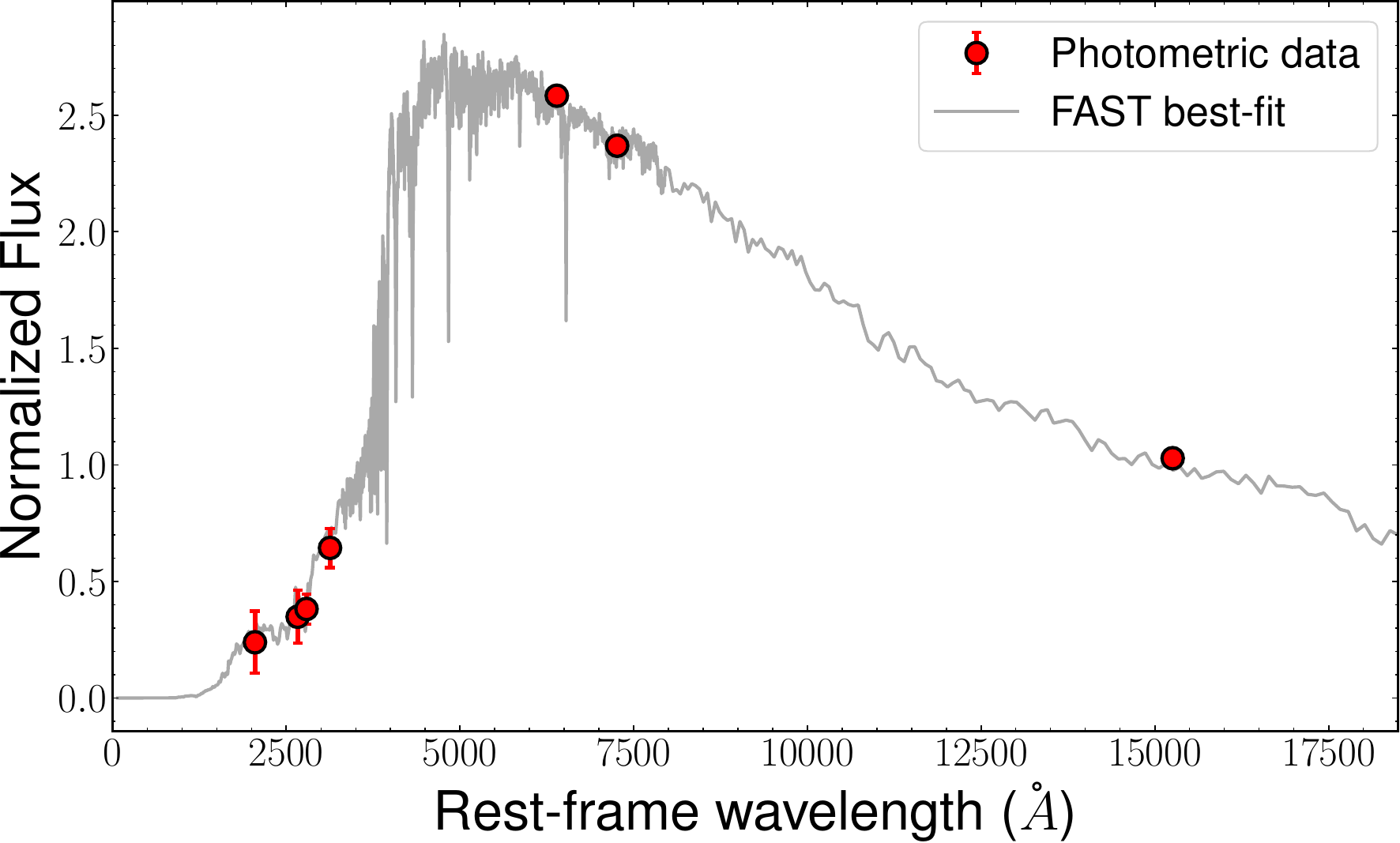}
\caption{FAST best-fit to the photometric data (red circles). Specifically, we used the convolved Kron magnitudes in the JWST filters F182M, F210M, and F444W, and the HST filters F606W, F775W, F814W, F850LP, as listed in the JADES photometric catalog.}
\label{fig:fastfit}
\end{figure}

To estimate the stellar mass of J-138717, we fitted the photometry provided by the JADES photometric catalog using the convolved Kron magnitudes. To fit the data, we used the C++ implementation of the FAST code \citep{fast}\footnote{Available from \url{https://github.com/cschreib/fastpp}} fed with the \citet{BC03} models using a Kroupa IMF and [M/H] = -0.4 dex. Further, we included a \citet{KC13} dust attenuation law in the fit. We estimate a stellar mass M$_* = 3.5\pm0.2 \times 10^{10}$ M$_\odot$\footnote{We find a consistent stellar mass using the FSPS models \citep{FSPS1}, namely M$_* = 3.3\pm0.2 \times 10^{10}$ M$_\odot$.}. The FAST best fit is shown in Fig. \ref{fig:fastfit}. FAST also estimates an age of 0.9 Gyr and attenuation parameter A$_V = 0.8$ mag, similar to the values obtained from the full-spectral fitting. 

We estimate the SFR from the H$\alpha$ and [OII] emission lines as follows. The fluxes reported in the previous section correspond to luminosities $L_{H\alpha} = 1.2\pm0.2\times 10^{40}$ erg s$^{-1} $ and $L_{[OII]} = 1.3\pm0.1 \times 10^{40}$ erg s$^{-1}$, given that $L = F \times 4 \pi d_L^2$, where $d_L = 1.44 \times 10^5$ Mpc is the luminosity distance at the redshift of the galaxy. We correct the observed luminosities for the dust attenuation as $L_{corr} = L_{obs} \times e^{\tau_\lambda}$, with $\tau_\lambda = A_\lambda/1.086$ and $A_\lambda = k_\lambda A_V/R_V$, where $A_V = 0.65$ mag, as estimated by the fit of the optical spectrum, $k_\lambda$ is the reddening curve for the gas, for which we assume a \citet{Calzetti+00} law with $R_V = 4.05$. Thus, $\tau_{H\alpha} \simeq 1.045$ and $\tau_{[OII]} \simeq 0.593$. We then calculate the SFR from H$\alpha$ using equation 2 of \citet{K98}, SFR$(H\alpha)$ = $7.9\times 10^{-42}\,L_{H\alpha, corr} = 0.16\pm0.03$ M$_\odot$ yr$^{-1}$, and from [OII] using equation 4 of \citet{Kewley+04}, SFR$([OII]) = 6.58\pm1.65\times 10^{-42}\,L_{[OII], corr} = 0.24\pm0.09$ M$_\odot$ yr$^{-1}$. Given the stellar mass estimated above, we derive the specific star formation rate (sSFR) as SFR(H$\alpha$)/M$_* = 5\pm1 \times10^{-12}$ yr$^{-1}$ and SFR([OII])/M$_* = 7\pm2 \times 10^{-12}$ yr$^{-1}$, both about 1 dex below the limit defined by \citet{Gallazzi+14}, i.e. sSFR$ < 1/5 t_{\rm U} (z) \simeq 6 \times 10^{-11}$, where $t_{\rm U} \simeq 3.4$ Gyr is the age of the Universe at the galaxy's redshift. This confirms the quiescent nature of J-138717. We note that emission lines usually suffer more dust attenuation than the stars. For local starburst galaxies, \citet{Calzetti+00} estimated that the color excess for stars is a factor 0.44 smaller than the gas (equation 3). This would imply an $A_V$ a factor $\sim2.3$ stronger for the gas than stars, such that we could underestimate the SFR by a factor $\sim2-4$ (depending on the line considered). However, even in that case, we would measure an sSFR lower than $ 1/5 t_{\rm U} (z)$. We also note that the resolution of the prism does not allow to resolve the blending between the H$\alpha$ and [NII] emission lines, so it is possible that the SFR derived from H$\alpha$ is underestimated. For this reason, we re-fit the prism spectrum without [NII], i.e. assuming that all the emission line is due to H$\alpha$. We measure $F(H\alpha) = 1.1\pm0.5\times 10^{-18} \mbox{erg cm}^{2} \mbox{s}^{-1} $, resulting in a SFR$(H\alpha) = 0.37\pm0.15$ and then sSFR $= 11\pm5 \times10^{-12}$ yr$^{-1}$, still consistent with the previous estimates. On the other hand, we note that, if J-138717 is a LINER (see Sect. \ref{sect:quenching}), then we are overestimating SFR(H$\alpha$), as the H$\alpha$ emission would be caused by an AGN or post-AGB stars, and not by the young stars. 

\begin{figure}
\includegraphics[width=\columnwidth]{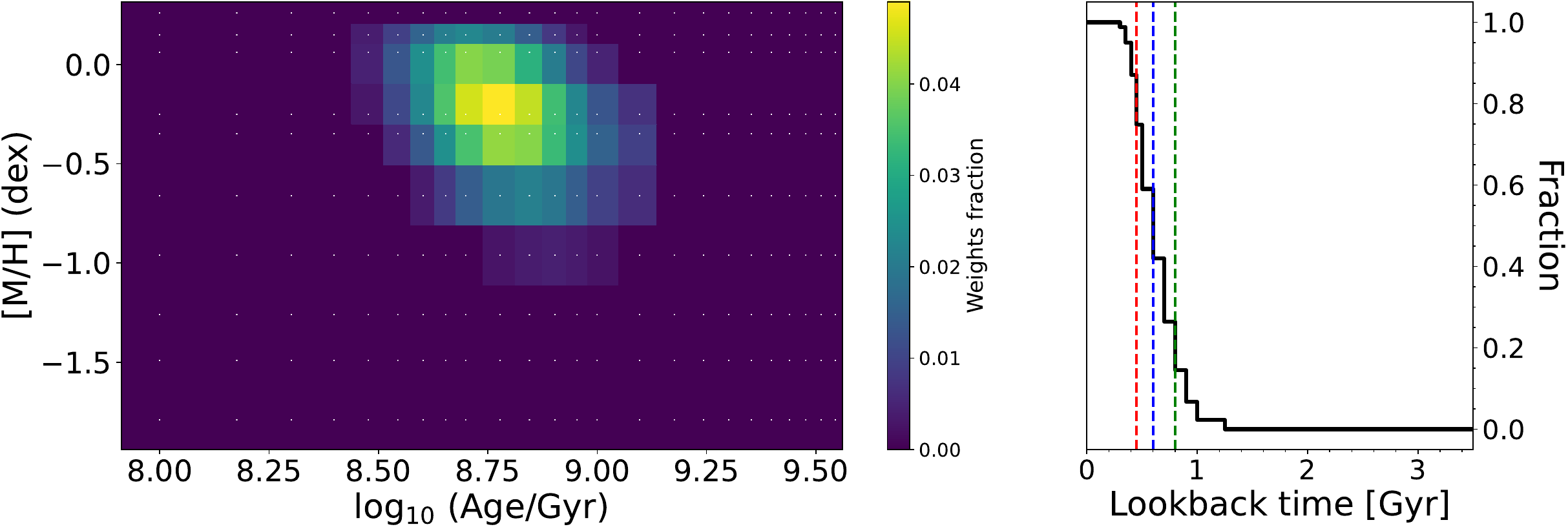}
\caption{SFH inferred from the regularized fit of the optical spectrum. Left panel: map of the weights assigned by \texttt{pPXF} to each input template of given age and metallicity. Right panel: cumulative sum of the weights. The green, blue, and red line correspond to $t_{20}^{\rm lbt}, t_{50}^{\rm lbt}, \mbox{and} \, t_{90}^{\rm lbt}$, respectively.}
\label{fig:sfh}
\end{figure}

We reconstruct the SFH of J-138717 in a non-parametric way, by performing the fit of the G140M spectrum with \texttt{pPXF} using the same setup described in the previous section but allowing for regularization \citep{ppxf_2017, ppxf_2023}. The SFH is shown in Fig. \ref{fig:sfh}. From the cumulative sum of the mass-weights, shown in the right panel, we estimate the lookback times at which the galaxy formed $20\%, 50\%, \mbox{and} \, 90\%$ of its mass, $t_{\rm 20}^{\rm lbt} = 0.8\pm0.1$ Gyr, $t_{\rm 50}^{\rm lbt} = 0.6\pm0.1$ Gyr, and $t_{\rm 90}^{\rm lbt} = 0.4\pm0.1$ Gyr, corresponding to redshifts $z_{20} \simeq 2.6$, $z_{50} \simeq 2.3$, and $z_{90} \simeq 2.1$, respectively.  Given $\Delta \tau = t_{90} - t_{20} \simeq 0.4$ Gyr, the average past SFR with which J-138717 formed its stellar mass is $\left \langle \mbox{SFR} \right\rangle = \mbox{M}_* / \Delta \tau \simeq 88$ M$_\odot$ yr$^{-1}$. We verified that fitting the NIR and prism spectra provides qualitatively the same SFH, but the lookback times are systematically older (for the NIR we get $t_{\rm 20}^{\rm lbt} = 1.2\pm0.2$ Gyr, $t_{\rm 50}^{\rm lbt} = 0.8\pm0.1$ Gyr, and $t_{\rm 90}^{\rm lbt} = 0.5\pm0.1$ Gyr) and the two $\Delta\tau$ are almost a factor 2 longer, resulting in a halved $\left \langle \mbox{SFR} \right\rangle$. This is in agreement with the older ($\sim 0.1$ Gyr) age found when performing the full-spectral fitting of the NIR compared to the optical (see Sect. \ref{sect:agemet}). This difference could arise because the Balmer lines are more sensitive to the most recent stellar burst (i.e. youngest population), while the NIR range is more sensitive to the older stellar component, thus hinting at the presence of a non-negligible fraction of an older ($\gtrsim 1$ Gyr) population. We caution that models in the NIR are uncertain at such young ages, as we show in the next section and further discuss in section \ref{sect:discussion}.

\section{Analysis of the spectral indices}\label{sect:indices}

\begin{figure}
\includegraphics[width=\columnwidth]{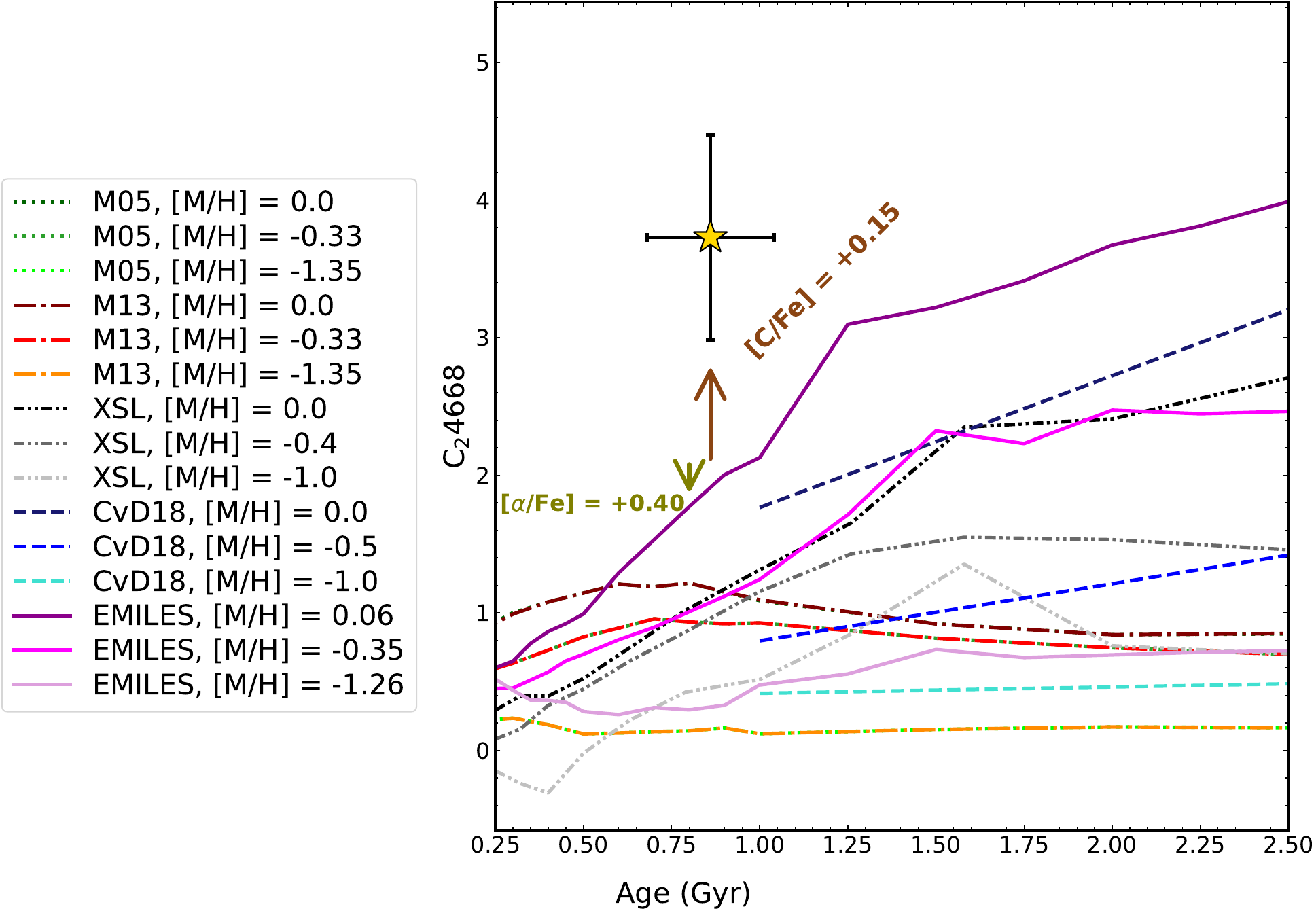}
\caption{C$_2$4668 spectral index as a function of age. The yellow star is the measured value from the G140M spectrum of J-138717. Green, red, grey, blue, and purple lines are the predictions of the M05, M13, XSL, CvD18, and EMILES models, respectively. Dark to light colors indicate higher to lower metallicities. Both observations and models have been smoothed to the instrumental resolution of the BaSeL stellar library. The brown (olive) line in the central (right) panel indicates by how much the tracks would shift towards higher index values due to a C-enhancement($\alpha$-enhancement) of 0.15 dex (0.4 dex), as predicted by the CvD18 (MILES) models.}
\label{fig:c2na}
\end{figure}

In this section, we compare the observations of the spectral indices with the predictions of the different models libraries, using the definitions of \citet{Trager+98} and \citet{eftekhari+21} for the optical and NIR indices, respectively. Considering the data reduction issues discussed in Sect. \ref{sect:data} (see also appendix \ref{app:DR3_B24_spec}), we study only those indices whose measurements are consistent when measured from the spectra of the DR3 and B24 data reductions. We also consider the optical spectral indices at rest-frame wavelengths $\lambda > 4400 \, \AA$, although the G140M spectrum of B24 does not reach longer wavelengths, so we can not evaluate the consistency of the measurements. Although consistent, some spectral indices are significantly different when measured from DR3 or B24. For this reason, we verified that the qualitative results of the following analysis do not change when considering the indices measured from B24. We found that the only index providing qualitatively different results is AlI1.67 (see section \ref{sect:nirindices}).

To compare observations with model predictions we have to match the different resolutions. Among the libraries considered in this work, the M05 and M13 models have the lowest resolution, which is also lower than that of the gratings, varying from $10 \, \AA - 100 \,\AA$ FWHM in the rest-frame wavelength range of J-138717. For this reason, we convolved the observed grating spectra and the EMILES, XSL, and CvD18 models by a Gaussian corresponding to the instrumental resolution of the M05 (and M13) models at the wavelengths corresponding to each spectral index. 

As the reference age we consider the one estimated from the full-spectral fitting of the optical spectrum using the EMILES models (sect. \ref{sect:agemet}), although for a proper comparison we should use the predictions from the different models. In practice, the qualitative results would not change for most of the libraries. The only exception is for the M13 models, which predict a systematically younger age ($\sim0.6$ Gyr; see Fig. \ref{fig:agemet_SSPs}) compared to all the other libraries. In many cases, M13 models predict values of the spectral indices closer to observations if an age of 0.6 Gyr is assumed. We highlight this consistency in the following sections, when needed.

\subsection{Probing the C-, Na-, and $\alpha$-enhancement}\label{sect:c2nad}

In Fig. \ref{fig:c2na} we compare the observed C$_2$ spectral index at 4668 $\AA$ with predictions of the different models. None of the models can reproduce the observations. Since this C$_2$ index is particularly sensitive to the C abundance, the observed mismatch could indicate that J-138717 is C-enhanced. To probe this, using the CvD18 model at 1 Gyr and solar metallicity, we calculate the difference in C$_2$ for a C-enhancement of [C/Fe] $= +0.15$ dex, shown in Fig. \ref{fig:c2na} with a brown arrow. Assuming C-enhancement significantly reduces the mismatch, but observations are still inconsistent with model predictions. This could imply a C-enhancement much higher than [C/Fe] $= +0.15$. This is in contrast with the recent results of \citet{Beverage+24}, who found a carbon deficiency in quiescent galaxies at comparable redshifts (although they also report a galaxy with a C-enhancement of +0.3 dex). We note, however, that C$_2$4668 is sensitive to the age (see the steep rise for the EMILES models in Fig. \ref{fig:c2na}), and the possible presence of a population of stars older than $\sim1$ Gyr, as the SFH indicates, could account for part of the mismatch. Finally, we verified that changing the IMF or [$\alpha$/Fe] has a negligible effect on this index at such a young age. As an example, in Fig. \ref{fig:c2na} we show the effect of increasing the [$\alpha$/Fe] from solar-scaled to +0.4 dex\footnote{This is estimated using the MILES models with varying $\alpha$-enhancements, assuming the age and metallicity estimated by full-spectral fitting of the optical range.}.

Similarly to C$_2$4668, models largely underpredict the NaD spectral index. We verified that the strength of the NaD index can not be recovered by any model even assuming a strong Na-enhancement, a steep IMF or $\alpha$-enhancement, and can only be explained by the presence of significant dust absorption, as we infer from the fits (Sect. \ref{sect:agemet}). In particular, we measure an equivalent width of $6.2\pm0.5 \, \AA$ for NaD, consistent with the measurements of \citet{Belli+24} for a PSB observed at $z=2.45$. Having $\mbox{EW}(\mbox{NaD}) = 6.5 \pm 0.5 \, \AA$, the authors estimate a hydrogen column density $N_H$\footnote{Calculated for the optically thin case, assuming a solar abundance value of (Na/H)$_\odot = -5.69$, the dust depletion index of the Milky Way, and a $10\%$ of Na being in the neutral phase.} of the order of $\sim10^{21}$ (see their equation 2). Following \citet{Rachford+09}, we estimate $N_H/E(B-V) = 5.8 \times 10^{21}$ H cm$^{-2}$mag$^{-1}$; assuming $R_V = A_V/E(B-V) = 3.1$ and substituting one gets $A_V \sim 0.7$ mag. Although this estimate is quite uncertain and relies on a number of assumptions, the result is consistent with the estimate of the full-spectral fitting and confirms the presence of a significant dust component. 

As shown in Fig. \ref{fig:fif}, J-138717 has a significant Mgb absorption that is not well matched by the EMILES models and could be indicative of a super-solar [$\alpha$/Fe] abundance. To probe $\alpha$-enhancement, we performed the FIF, as described in section \ref{sect:agemet}, fitting [$\alpha$/Fe] with the MILES models having [$\alpha$/Fe] =$+0.0$ and $+0.4$. From this fit, we estimate [$\alpha$/Fe] $= 0.15\pm0.10$ dex, hinting at $\alpha$-enhancement. Using a more traditional approach (e.g., \citealt{Thomas+03, Bevacqua+23}), we constructed grids of Mgb vs. Fe3, with Fe3 $= \frac{1}{3}\left(\mbox{Fe4383} + \mbox{Fe5270} + \mbox{Fe5335}\right)$ by interpolating the MILES models at different [$\alpha$/Fe] and metallicity, having a fixed age of 0.8 Gyr. We then estimate [$\alpha$/Fe] by minimizing $\chi^2$, defined as

\begin{equation}\label{eq:chi2}
\chi^2 = \sum_j \left(\frac{\rm{I}_{\rm{obs},\textit{j}} - \rm{I}_{\rm {mod},\textit{j}}}{\sigma_{\rm{I},\textit{j}}}\right)^2 \; \; \; ,
\end{equation}
\noindent where j runs over the indices Mgb and Fe3, I$_{\rm{obs}, \textit{j}}$ are the observed indices, I$_{\rm{mod}, \textit{j}}$ are the indices of the model grid, and $\sigma_{\rm {I},\textit{j}}$ is the measured error. With this method, we estimate [$\alpha$/Fe] $= 0.19\pm0.13$. Both the FIF and model grid suggest that J138717 might be $\alpha$-enhanced, but the uncertainties are large and a higher SNR is required to safely constrain [$\alpha$/Fe].

In summary, the analysis of C$_2$4668 suggests that J-138717 is strongly C-enhanced ([C/Fe]$>0.15$ dex), but the actual enhancement is questioned by the possible presence of an old ($>1$ Gyr) stellar component affecting the index. The analysis of NaD does not allow us to estimate the [Na/Fe] abundance, since the strength of the index is dominated by the dust. Assuming that the measured EW(NaD) is determined by the dust absorption, we roughly estimate A$_V \sim 0.7$ in agreement with the fits. Finally, the Mgb and iron lines indicate that J-138717 is $\alpha$-enhanced ([$\alpha$/Fe]$\sim0.15$ dex), but the SNR is too low to put a stringent constrain on [$\alpha$/Fe]. 

\begin{figure*}
\includegraphics[width=\textwidth]{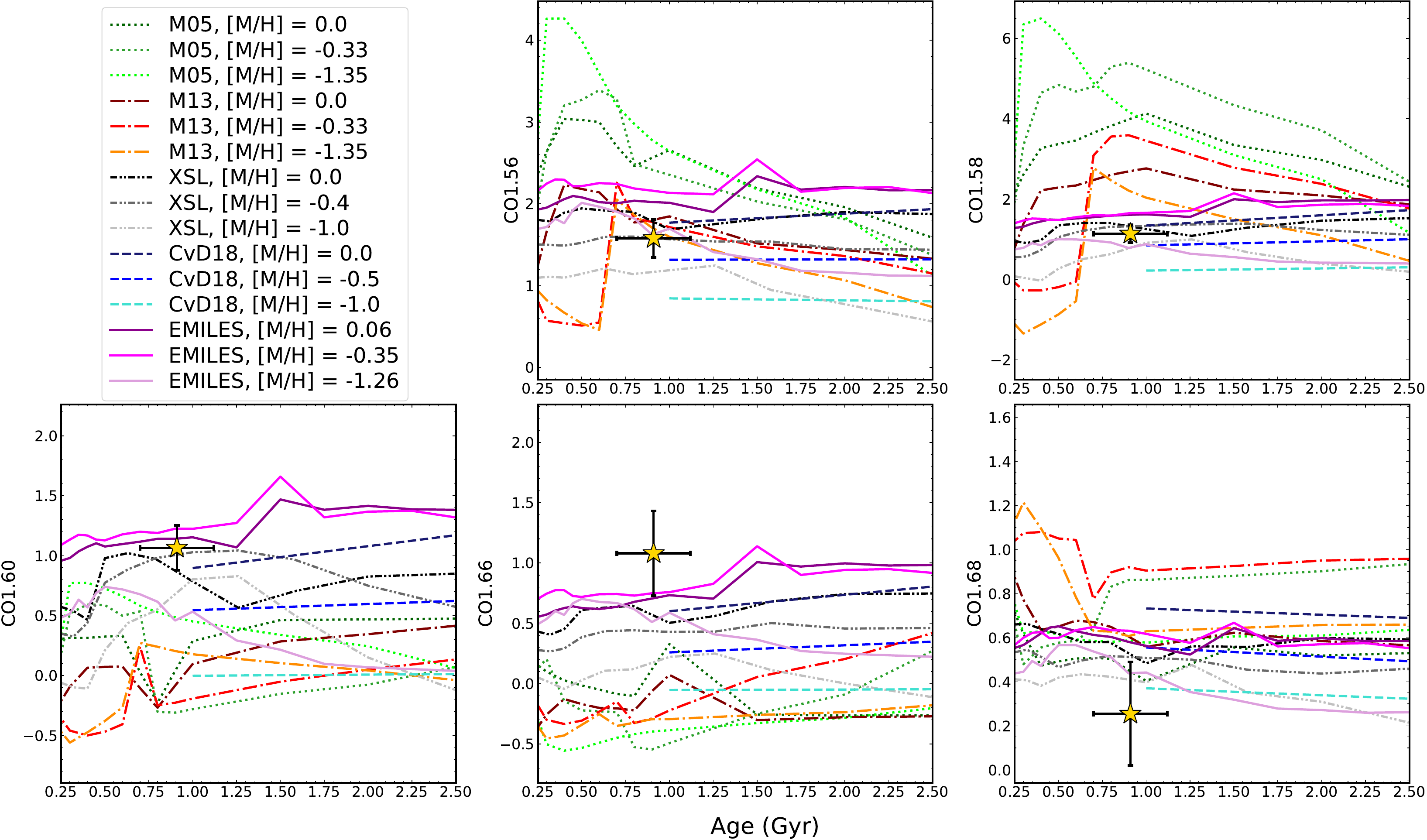}
\caption{CO indices as a function of age. The marker and the lines are the same as in Fig. \ref{fig:c2na}}
\label{fig:CO_indices}
\end{figure*}

\begin{figure*}
\includegraphics[width=\textwidth]{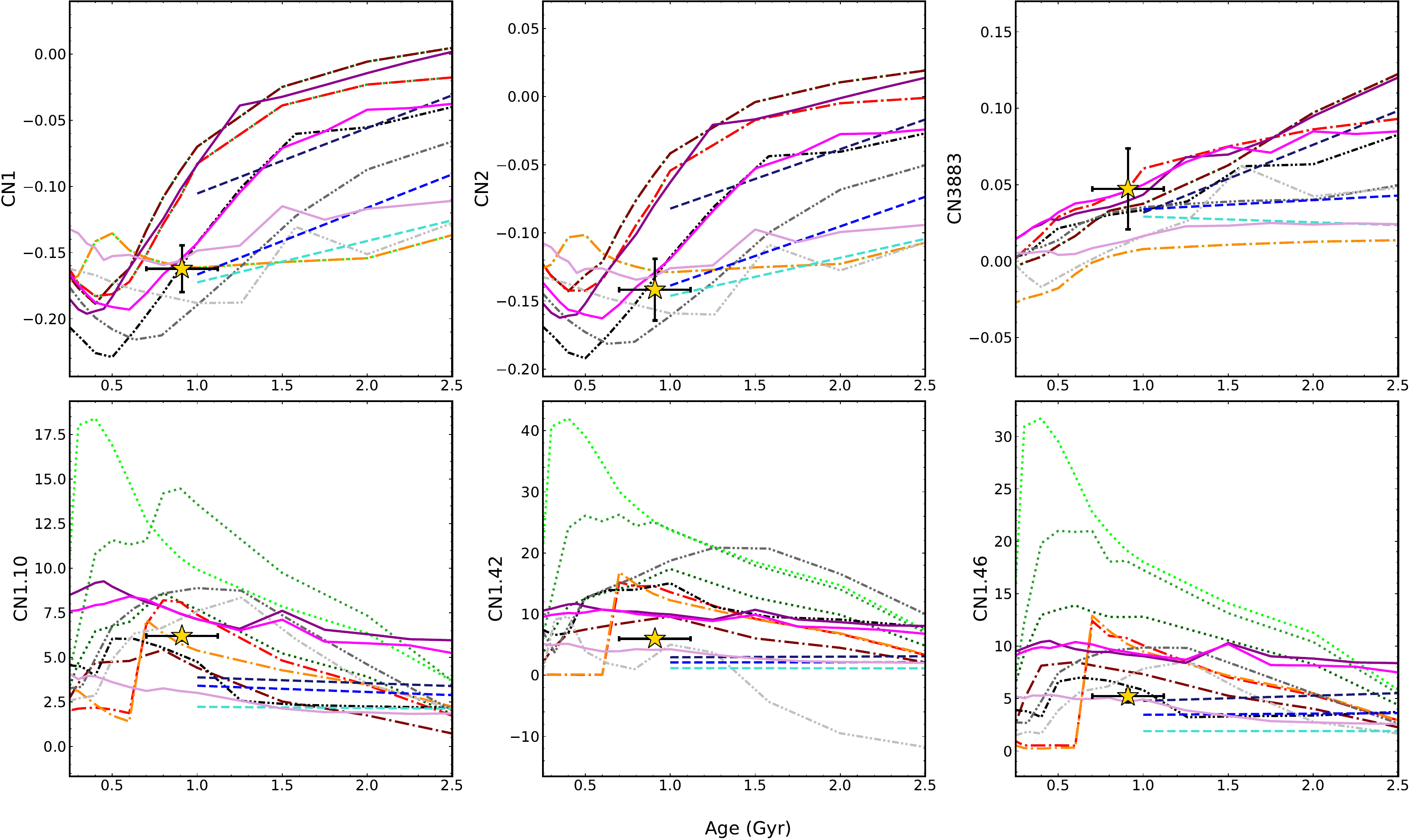}
\caption{CN indices as a function of age. The marker and the lines are the same as in Fig. \ref{fig:c2na}.}
\label{fig:CN_indices}
\end{figure*}

\begin{figure*}
\includegraphics[width=\textwidth]{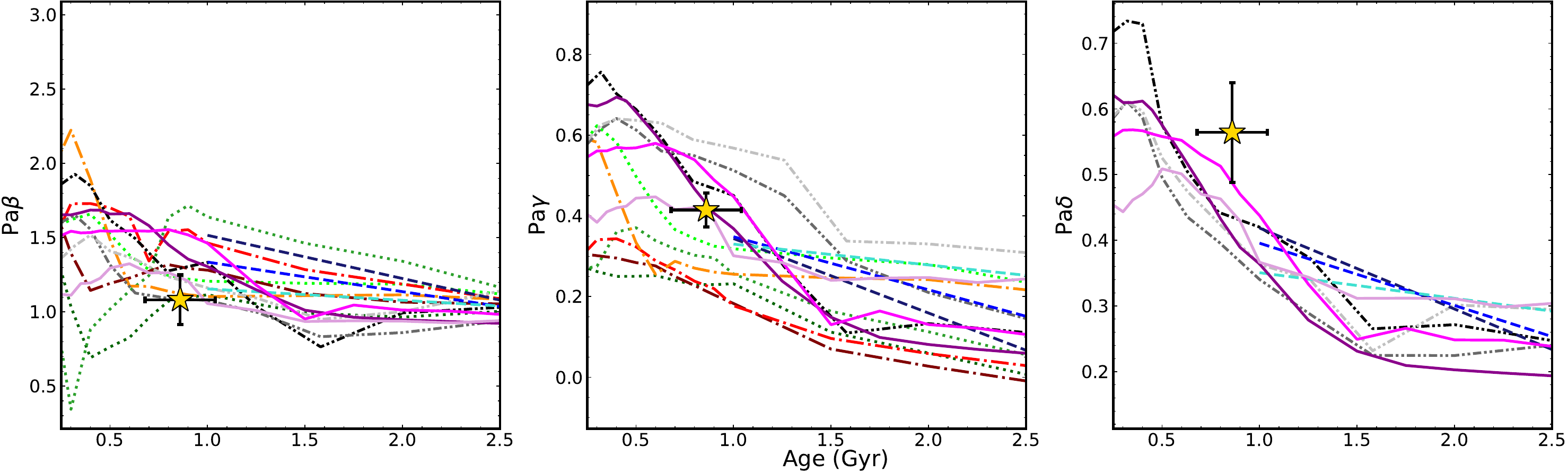}
\caption{Paschen series indices as a function of age. The marker and the lines are the same as in Fig. \ref{fig:c2na}. Predictions from M05 and M13 for Pa$\delta$ are not shown because this index can not be measured safely due to the low spectral resolution (see text for details).}
\label{fig:Pa_indices}
\end{figure*}

\begin{figure*}
\includegraphics[width=\textwidth]{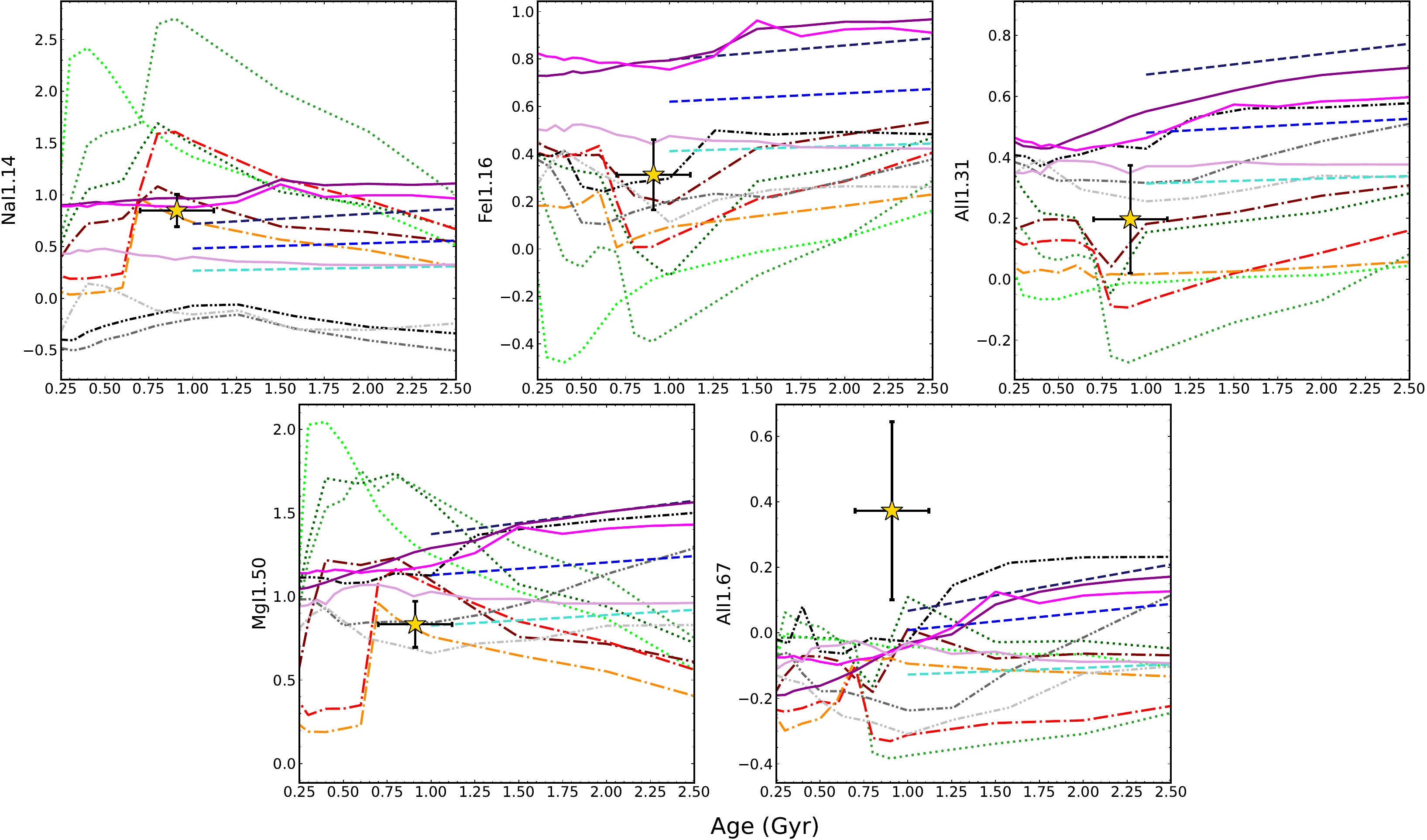}
\caption{Other NIR indices as a function of age. The marker and the lines are the same as in Fig. \ref{fig:c2na}.}
\label{fig:NIR_indices}
\end{figure*}

\subsection{Probing the contribution of TP-AGB stars}\label{sect:cocn}

The most sensitive spectral indices to TP-AGB stars are the molecular CO and CN bands \citep{M05, Riffel+07, Martins+13}. In Fig. \ref{fig:CO_indices} and \ref{fig:CN_indices} we show these indices as observed for J-138717 and compared to the model predictions. We recall that M05 and M13 are semi-empirical models accounting for a heavy and mild contribution of TP-AGB stars, while EMILES, XSL, and CvD18 are based on fully empirical libraries and do not have a significant contribution. 

In Fig. \ref{fig:CO_indices} we show the CO1.56, CO1.58, CO1.60, CO1.64, and CO1.68 indices. In general, none of the libraries can simultaneously match all the observed CO indices. Models fully based on empirical  stellar libraries (EMILES, XSL, and CvD18) generally provide more consistent measurements with observations. However, the inferred metallicities vary significantly from index to index and from library to library. For instance, the EMILES and XSL models predict a subsolar metallicity consistent with that estimated from full-spectral fitting for CO1.58 and CO1.60, while CO1.56 and CO1.68 suggest much lower metallicity, and finally, CO1.66 requires a supersolar metallicity to match the observations. The M13 models would be consistent with the observations only for the CO1.56 and CO1.58 index, assuming an age of $\sim 0.6$ Gyr,  as predicted by the fits, while the other predictions do not reproduce the observations. Finally, the M05 models do not match any of the observed CO indices. We verified that the M05 and M13 models could match all the observed CO indices only when assuming a very high metallicity ($>0.35$ dex), but this is in sharp contrast to the metallicity estimated from the fits, including those obtained with the M05 and M13 models (Fig. \ref{fig:agemet_SSPs}).

In Fig. \ref{fig:CN_indices} we show the CN1, CN2, CN3883, CN1.10, CN1.42, and CN1.46 indices. In this case, all libraries can match the optical CN indices (CN1, CN2, and CN3883). The EMILES models can also reproduce all the observed NIR indices of the CN with subsolar metallicities (although CN1.42 and CN1.46 require metallicities much lower than that estimated from the fits). The XSL models can reproduce the CN1.10 and CN1.42 (and only approximately CN1.46) indices, but the inferred metallicity again depends on the index considered. The CvD18 models tend to underpredict all the NIR CN indices and they would require supersolar metallicities to match the observations. The M13 models are consistent only with the CN1.10 in the NIR, but they would be consistent with all CN indices assuming an age of 0.6 Gyr. Finally, the M05 model predictions do not match any of the NIR CN indices. Similarly to the CO indices, the M05 and M13 models could match the CN indices assuming a very high metallicity.

For all CO and CN indices, we also checked the effect of C-enhancement, $\alpha$-enhancement, and IMF variation and found contrasting results. For instance, a C-enhancement would decrease both the CO1.66 and CO1.68 predicted by the models, resulting in a worse match for the former and a better match for the latter when compared to observations. Similar results are found for the other indices and a similar effect is obtained when varying the IMF, while varying [$\alpha$/Fe] has a negligible effect in all cases. Finally, we note that \citet{FLB+24} verified that the use of different isochrones does not significantly affect the values of the indices measured from the models. 

Overall, none of the libraries can simultaneously match all the CO and CN observed indices. The models based on empirical stellar libraries (EMILES, XSL, and CvD18) generally predict CO and CN indices closer to observations than the theoretical ones (M05 and M13), but the inferred metallicity can vary significantly from index to index and from library to library. Instead, the theoretical models struggle to reproduce the observations, and in particular the NIR indices. The M13 models provide predictions consistent with the observations of all CN indices and two CO indices assuming an age of $\sim 0.6$ Gyr, consistent with the estimates from full-spectral fitting, but systematically lower than the estimates of all the other libraries. Both the M05 and M13 models could match observations assuming a very high metallicity, but this is in contrast with the metallicity estimates from the fits. In conclusion, our analysis indicates that there is no heavy contribution of TP-AGB stars to the spectrum of J-138717.

\subsection{Other NIR spectral indices}\label{sect:nirindices}

In Fig. \ref{fig:Pa_indices} we show the comparison of the measured and predicted Paschen lines, which are age indicators of a galaxy, as they do not depend significantly on the metallicity. The Pa$\beta$ line is consistent with all libraries. The Pa$\gamma$ is in agreement with the EMILES, XSL, and CvD18 models. Instead, the M05 and M13 models underpredict this index. Similarly to the CN indices, considering an age of 0.6 Gyr would provide consistent predictions for the M13 (and M05) models. Finally, none of the libraries can reproduce the Pa$\delta$ line (barely, the EMILES models with solar metallicity). We note that the Pa$\delta$ predictions for the M05 and M13 models are not shown because, due to the large wavelength sampling of the BaSeL library from which they are constructed, there are only two spectral pixels sampling the Pa$\delta$ bandpass, resulting in an unrealistic flat shape of the index.

In Fig. \ref{fig:NIR_indices} we show other NIR spectral indices. Overall, none of the libraries can match all these indices. In general, the M13 models reproduce most of these indices, but the inferred metallicity changes significantly from index to index. The NaI1.14 and FeI1.16 indices are consistent with all libraries, except the M05 models. The AlI1.31 index is consistent with all libraries, but the EMILES (M05 and M13) models require very low (high) metallicities. The MgI1.50 index is consistent with the M13, XSL, and CvD18 libraries that have different metallicities. Finally, the AlI1.67 index is inconsistent with all libraries. However, when measured from the B24 spectrum, although it is consistent with the measurements from the DR3 spectrum (given the large uncertainty), it is much lower and consistent with all libraries. We conclude that the modeling of the NIR spectral indices at young ages is uncertain for all of the state-of-the-art models considered in this work.

\section{Discussion}\label{sect:discussion}

\subsection{SFH and quenching of J-138717}\label{sect:quenching}
\begin{figure}
\includegraphics[width = \columnwidth]{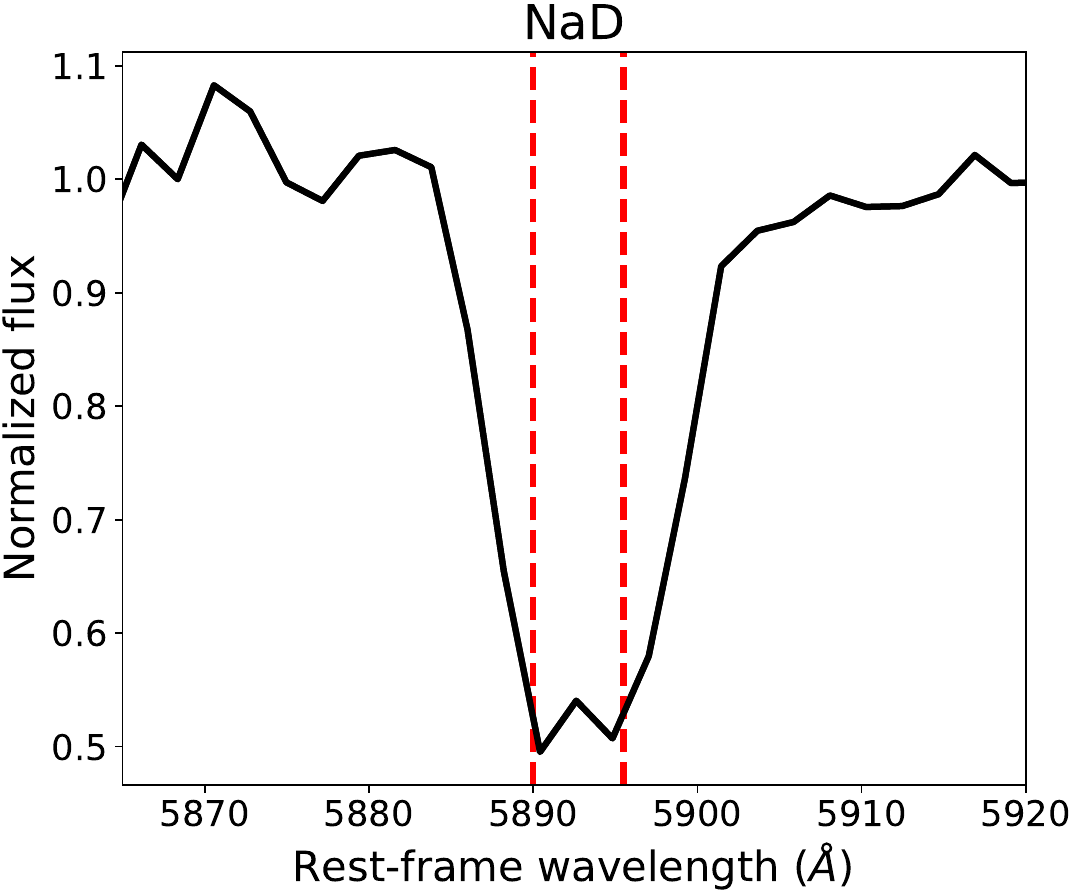}
\caption{Observed profile of the NaD absorption line (black line). The red dashed lines indicate the expected wavelengths of the doublet at $\lambda = 5890 \, \AA$ and $\lambda = 5896 \, \AA$.}
\label{fig:nad}
\end{figure}

The SFH shown in Fig. \ref{fig:sfh} indicates that J-138717 formed most of its mass ($\geq50\%$) recently ($t_{50}^{\rm lbt} = 0.6$ Gyr) and then quenched recently and rapidly ($t_{90}^{\rm lbt} = 0.4$ Gyr). The mechanisms of rapid quenching are generally attributed to gas-rich mergers and strong AGN feedback. \citet{Park+24} recently found that the majority of rapidly formed and recently quenched galaxies, such as J-138717, often exhibit strong AGN features (\citealt{Davies+24}; see also \citealt{Taylor+24}). The spectrum of J-138717 indeed shows high-ionization lines typical of AGNs, namely [NeIII], [OIII], and [NII]. However, the relatively low flux ratios log$_{10}$([OIII]/[OII]) $\simeq -0.1$ dex and log$_{10}$([NeIII]/[OII]) $\simeq -0.5$ dex suggest that these emissions are sustained by some low level star formation. If this is the case, such residual star formation would be consistent with a scenario where stars formed very efficiently, quickly consuming most of the cold gas and then halt (\citealt{Vazdekis+96, Vazdekis+97, Weidner+13, Jerabkova+18, SR+20}), which could also explain the compact size of J-138717 ($\simeq 1.1$ kpc).

On the other hand, the relatively high flux ratio log$_{10}$([NII]/H$\alpha) \simeq 0.5$ dex suggests that J-138717 is a LINER galaxy \citep{Kewley+06}\footnote{Please, note that the [NII] and H$\alpha$ lines are blended in the prism spectrum, so the flux ratio we measured could be overestimated.}. This would be in agreement with other studies showing that many PSBs have LINER emissions (e.g., \citealt{French+15}), as also recently found in some other massive quiescent galaxies observed at comparable redshifts by \citet{Bugiani+24}. The LINER emission we detect could be due to weak AGN activity, possibly preventing new star formation over the past 400 Myr, although post-AGB stars and shocks can also be responsible for the LINER emissions \citep{Belfiore+16, Yan18}. 

Quenching via strong AGN feedback in massive galaxies at high redshifts is often associated with massive multiphase outflows \citep{Belli+24, Davies+24, Carniani+24, Park+24, Taylor+24}. These outflows are identified in the galaxy spectrum by a shift of the emission and absorption lines tracking the gas with respect to their systemic velocities or by asymmetric line profiles. For J-138717, the emission lines do not exhibit any asymmetry or shift in wavelength, as shown in Fig. \ref{fig:emlines}. Then, in Fig. \ref{fig:nad} we show the profile of the NaD absorption line, tracking the neutral gas. The observed lines are centered at the expected wavelengths, their shapes are Gaussian and symmetric, and their line strengths are comparable. We verified that the CaII (H\&K) lines (shown in Fig. \ref{fig:specind_dr3_b24}) and MgII doublet at $\lambda~2796, 2804 \, \AA$, two other tracer of the neutral gas associated with outflows \citep{Nestor+11, Belli+24, Taylor+24}, show similar, well-behaved profiles.
 
Overall, our analysis suggests that there is no strong AGN feedback in J-138717. However, since the AGN emission is not typically expected to be spherically symmetric, we could be missing this evidence due to the direction of the AGN emission and outflows being misaligned with respect to the observations. Other mechanisms associated with rapid quenching are wet mergers and gas-stripping. Both these cases would require a study of the environment of J-138717. Although we do not find any sources in its proximity at comparable redshift (the closest source among galaxies with $1.8 < z_{\rm phot} < 2.3$ is $\sim 14''$ away, corresponding to $\sim 110$ kpc), we do not attempt any detailed environmental study of J-138717, because no spectroscopic redshifts are available for close-by sources and the uncertainties on the photometric redshifts are too large (for example we measure a difference $z_{\rm phot} - z_{\rm spec} \sim 0.3$ for J-138717).

\subsection{The low contribution of TP-AGB stars}\label{sect:tp-agb}

According to the predictions of the fuel-consumption theorem \citep{RV81}, TP-AGB stars are expected to dominate the luminosity of an SSP at young ages ($0.2-2$ Gyr), accounting for up to $80\%$ of the NIR and $40\%$ of the total luminosity \citep{M98}. The JWST observations of J-138717 offer a unique opportunity to probe the actual contribution of TP-AGB stars, as the resolution of the NIRSpec gratings allows for reliable measurements of the NIR spectral indices, and, given the young age and short SFH of this post-starburst galaxy, the contribution of TP-AGB stars should be at its peak. We highlight that such analysis has never been done before outside the local Universe. From our analysis, models based on empirical stellar libraries (EMILES, XSL, and CvD18), having a low contribution of TP-AGB stars, can reproduce a large number of CO and CN indices, tracing TP-AGB stars. The M13 models, having a mild contribution of TP-AGB stars, can reproduce many CO indices and CN indices assuming an age of $\sim0.6$ Gyr, which would be consistent with the fits (Fig. \ref{fig:agemet_SSPs}). A younger age, combined with some residual star formation, could indeed dilute the signal of TP-AGB stars. On the other hand, we note that all the other models predict systematically older ages. Finally, the M05 models, having a heavy contribution of TP-AGB stars, do not reproduce any of the observed CO and NIR CN indices. These results suggest that the contribution of TP-AGB stars to the spectrum of J-138717 is not significant, in agreement with other PSBs studied by \citet{Kriek+10} at high redshifts and \citet{CG10} in the local Universe.

We note that \citet{eftekhari_CO}, found similar results for local massive ETGs, but we find a lower mismatch between observations and model predictions for the CO and CN indices. This might indicate that the TP-AGB contribution varies (or the modeling of the CO and CN features gets worse) at older ages, higher metallicities, and, possibly, steeper IMF slopes that characterize local massive quiescent ETGs. For example, \citet{MG+21} argued that the luminosity of TP-AGB stars may vary with metallicity as it controls their mass-loss rate. Overall, an improvement in the modeling of the NIR spectrum is still very much needed. \citet{Riffel+15} suggested that stars in advanced evolutionary phases, like red giant branch (RGB) and early-AGB stars, may help reduce the observed mismatch. We further discuss this issue in the next section.

Recently, \citet{Lu+24} reported the presence of strong absorption features, such as CN and CO, in the prism spectra of three young quiescent galaxies found at redshifts $z=1-2$, similarly to J-138717, and attributed these features to a significant contribution of TP-AGB stars. Unfortunately, the resolution of the prism does not allow us to measure the NIR spectral indices. We have shown that, although J-138717 also shows strong CO and CN absorptions detectable in the prism data, the spectral indices analysis disfavors a heavy contribution of TP-AGB stars. Therefore, observations with higher resolution and SNR are needed to confirm the results found by \citet{Lu+24}. It is worth noting that the galaxy spectra shown by \citet{Lu+24} exhibit stronger CO, CN, and C$_2$ features than J-138717 as well as other features (such as TiO, VO, and ZrO) that are not detected in the spectrum of J-138717. Although this can be explained by the presence of other poorly modeled stars, like carbon-rich, early AGB, and RGB stars \citep{Riffel+15}, they are indeed three valuable candidates to further constrain the contribution of TP-AGB stars in galaxies. If confirmed, a heavy contribution of TP-AGB stars in these three galaxies, at variance with J-138717, would imply that the significance of TP-AGB stars varies among galaxies. This would explain the conflicting results found in the literature and would call for considerable improvements in the modeling of evolved stars such as the TP-AGB.

\subsection{General issues and caveats}\label{sect:caveats}

This work offers the opportunity to discuss some issues related to the current data analysis and recent results of the literature. A major problem concerns the reliability of the data reduction pipeline of JWST. Although great endeavors have been made to provide reliable spectra, such as for the reduction of the JADES spectra (see \citealt{DR3}), some important problems still persist and can affect significantly the results. For example, we showed that there is a systematic offset between the prism and G395M spectrum (Fig. \ref{fig:spec}) that increases at longer wavelengths. This has been also reported in the DR3 of JADES. By comparing with B24, we showed that different data reductions provide significantly different spectra and thus different results. For example, the slope of the prism spectra of the DR3 is steeper than that of B24 (Fig. \ref{fig:dr3_b24}), resulting in a younger age differing up to 0.5 Gyr (depending on the models). In both cases, we could not fit the wavelength region $\lambda \sim 5000-6000 \, \AA$. Analogously, \citet{Lu+24} found that they can not fit the spectral region between $\lambda \sim 5000-10,000 \, \AA$, and ascribed this issue to the poor modeling of TP-AGB stars in state-of-art models. This does not seem to be the case for J-138717, since our analysis based on the spectral indices still disfavors a strong presence of TP-AGB stars. Most importantly, the G140M spectrum, covering this wavelength range, does not suffer from this problem and models can easily fit the continuum. Given the consistency of the G140M and prism spectra, and the mismatch between the G395M and prism spectra (see Fig. \ref{fig:spec}), it is likely that the flux calibration of the prism is incorrect at longer wavelengths. Such systematics can have a huge impact, especially when working with young galaxies at very high redshifts such as JADES-GS-z14-0 and JADES-GS-z14-1 \citep{Carniani_z14}, as two extreme examples. So far, different teams have adopted their customized (private) pipelines, but an effort should be made by the community to provide a reliable public data reduction pipeline. 

Another issue, concerns the reliability of the stellar population models. In this work, we have derived stellar population parameters with many widely used models\footnote{We here mention that we obtained similar results with other commonly used libraries, such as \citet{BC03} and FSPS \citep{FSPS1}, even though we do not show these results.}. We found a generally good agreement in the estimated parameters when performing the full-spectral fitting, although the parameters have large uncertainties notwithstanding the relatively high SNR of the spectra. Instead, the analysis of the NIR spectral indices (see Sect. \ref{sect:indices}) reveals major inconsistencies, as none of the models adopted in this work can reproduce all the observations, and those indices that are well reproduced by models provide inconsistent metallicities. For example, models based on empirical libraries predict better CO, CN, and Pa indices (Figs. \ref{fig:CO_indices}, \ref{fig:CN_indices}, \ref{fig:Pa_indices}), while M13 models better predict NaI, FeI, AlI, and MgI NIR absorption lines (Fig. \ref{fig:NIR_indices}), but the metallicity derived varies significantly from index to index and from library to library. The inconsistency of the metallicity derived from the NIR indices was already pointed out by \citet{Riffel+15}, who suggested that an improved treatment of TP-AGB stars, as well as red giants and early-AGB stars, might minimize the mismatch. We also note that the resolution of the M05 and M13 models (or other models in the literature using low-resolution NIR libraries) is potentially a source of major limitations in our data analysis (e.g., \citealt{DH+18}). For example, we have discussed how some indices, like Pa$\delta$, can not even be measured because the FWHM is larger than the absorption line. An effort to construct models with higher resolution (in particular, those including a significant contribution of TP-AGB stars) is invaluable for future studies based on NIR spectral indices. Overall, current stellar population models should be used with caution when deriving stellar population properties (e.g., the metallicity) from the NIR, in particular at young ages.

Many recent works in the literature are discussing results based on the SFHs of high-redshift galaxies, having profound implications on our understanding of galaxy formation. However, the recovery of the SFH is very challenging as it suffers intrinsic uncertainties that are difficult to overcome (e.g., \citealt{Zibetti+24}) and may depend on the methods adopted. For example, \citet{Glazebrook+24} found that ZF-UDS-7329, a massive (M$_* = 10^{11.4}$ M$_\odot$) quiescent galaxy observed at $z\simeq3.2$ that formed most of its mass already at $z=11$, as also confirmed by \citet{Carnall+24}. However, \citet{Crispin+24} showed that the actual formation redshift can be lowered to $z=7$ (i.e. doubling the age of the Universe), and the age can be by up to 0.5 Gyr younger, just by changing priors. In this work, we have shown that the spectral range fitted can also provide differences of a few hundred Myr in the SFH. For J-138717, this could be due to the presence of a non-negligible fraction ($\geq 20\%$) of old ($>1.2$ Gyr) stars detected in the NIR spectrum, but, considering the analysis of the NIR spectral features, we can not exclude that this is due to the difficulty of state-of-the-art models in reproducing the NIR spectrum. In the former case, this effect should be reduced at higher redshifts, when the stellar populations are homogeneously younger by construction. We note that \citet{Nanayakkara+25}, except for some galaxies, found statistically consistent SFHs when fitting different stellar population models or spectral ranges. However, we point out that the timescales they inferred from the different methods can have differences of the order of a few hundred Myr, in agreement with our results. Considering all these uncertainties, more compelling evidence is required to shake the foundations of the $\Lambda$CDM.

\section{Summary and conclusions}\label{sect:summary}

We studied the physical and stellar population properties of J-138717, a post-starburst galaxy observed at $z_{\rm spec} = 1.8845$ with JWST, for which we estimated a stellar mass M$_* = 3.5\pm0.2 \times 10^{10}$ M$_\odot$ and velocity dispersion $\sigma_* = 198\pm10$ km s$^{-1}$. We summarize our results as follows.

\begin{itemize}

\item From the full-spectral fitting of the optical, NIR, and prism spectra, we estimated an age of $0.91\pm0.21$, $0.95\pm0.25$, and $0.89\pm0.11$ Gyr, respectively, and a metallicity of $-0.28\pm0.17$, $-0.38\pm0.15$, and $-0.35\pm0.08$ dex, using the EMILES models. The FIF method fitting the optical spectral indices and fed with the EMILES models provides an age of $0.81^{+0.05}_{-0.07}$ Gyr and a metallicity of $-0.19^{+0.11}_{-0.07}$, consistent with the full-spectral fitting. We verified that using other models generally provides consistent results. 

\item From the optical and prism spectra, we also estimated dust attenuation parameters A$_V = 0.65\pm0.18 \, \mbox{and} \, 0.63\pm0.05$ mag, and $\delta = -0.82 \pm 0.29 \, \mbox{and} \, -0.90 \pm 0.12$ respectively. From EW(NaD) we also estimated A$_V \sim 0.7$, consistent with the fits. This is in agreement with other results in the literature reporting a significant dust component in high-redshift quiescent galaxies.

\item From the H$\alpha$ and [OII] emission lines, respectively, we estimated SFR$=0.16\pm0.03$ M$_\odot$ yr$^{-1}$ and $=0.24\pm0.09$ M$_\odot$ yr$^{-1}$, implying sSFR$= 5\pm1 \times 10^{-12}$ yr$^{-1}$ and sSFR$= 7\pm1 \times 10^{-12}$ yr$^{-1}$, about 1 order of magnitude lower than a threshold of $1/5 t_{\rm U} (z)$ to define quiescent galaxies.

\item We reconstructed the SFH by fitting the optical spectrum and found that J-138717 started forming its stars at $t_{20}^{\rm lbt} = 0.8\pm0.1$ Gyr, then built half of its stellar mass by $t_{50}^{\rm lbt} = 0.6\pm0.1$ Gyr, and quenched at $t_{90}^{\rm lbt} = 0.4\pm0.1$ Gyr. This implies a formation timescale of about $\Delta \tau = 0.4$ Gyr, corresponding to an average $\left\langle \mbox{SFR} \right\rangle = 88$ M$_\odot$ yr$^{-1}$. However, the fits of the NIR spectrum provides earlier $t_{20}^{\rm lbt} = 1.2\pm0.2$ Gyr, possibly suggesting the presence of an older population, that would imply a larger $\Delta\tau$ and a lower $\left\langle \mbox{SFR} \right\rangle$.

\item The low flux ratios $\log_{10}$([OIII]/[OII]$)\simeq -0.1$ dex and $\log_{10}$([NeIII]/[OII]$) \simeq -0.5$ dex suggest that these emissions are due to some residual, low-level star formation. The $\log_{10}$([NII]/H$\alpha) \simeq 0.5$ dex indicates LINER emission, which could be either due to a weak AGN or post-AGB stars. Furthermore, the symmetry of the emission and absorption lines observed at their systemic velocity argue against the presence of ongoing outflows. 

\item The analysis of the C$_2$4668 index hints that J-138717 is C-enhanced, but the possible presence of a stellar population older than 1 Gyr is degenerate with it. From the FIF and the Mgb-Fe3 model grids, we derived [$\alpha$/Fe] $= 0.15\pm0.10$ and $0.19\pm0.13$ dex, respectively, suggesting that J-138717 is $\alpha$-enhanced, but a higher SNR is needed to confirm this result. 

\item We measured the CO and CN absorption lines, tracing TP-AGB stars, and compared them with different model predictions. We found that the EMILES, XSL, and CvD18 models, based on empirical stellar libraries and having a low contribution of TP-AGB stars, match a large number of these indices. The M13 models, having a mild contribution of TP-AGB stars, match several indices assuming a younger age (0.6 Gyr), consistent with the fits. Finally, the M05 models, having a heavy contribution of TP-AGB stars, do not match any of the CO or CN indices. These results indicate that the contribution of TP-AGB stars to the spectrum of J-138717 is not significant.

\item We measured several other NIR indices and compared them with model predictions. We found that the M13 models can match almost all of them. The EMILES, XSL, and CvD18 also reproduce many of those indices. In all cases, we inferred inconsistent metallicities, varying significantly from index to index and from library to library. Finally, the M05 models are consistent only with a couple of NIR spectral indices. We highlight that this is the first time that such a detailed analysis of the NIR spectral indices is performed outside the local Universe.

\end{itemize}

The SFH suggests that J-138717 formed most of its mass quickly and then quenched rapidly, but with current data it is not possible to clearly determine what caused its quenching. The analysis of the CO and CN spectral indices suggests that TP-AGB stars do not account for a significant fraction of stellar luminosity, contrary to what is expected by theoretical predictions for young PSBs such as J-138717. 

In general, it is reassuring to find consistent estimates  of the stellar population properties from the full-spectral fitting when fitting different wavelength ranges and using different model libraries. On the other hand, all models struggle to reproduce the NIR spectral indices consistently, pointing to the importance of improving state-of-the-art stellar population models in the NIR, especially at young ages and low metallicities, which is most relevant to studying high-redshift galaxies in the JWST era. \\

\textit{Data availability.} All the data from the DR3 are publicly available and can be downloaded from the MAST archive at \url{https://archive.stsci.edu/hlsp/jades}. The B24 spectra are publicly available and can be retrieved from the Dawn JWST Archive. \\

\begin{acknowledgements}
We thank the referee for the constructive report. D.B. acknowledges support by the Archival Research Visitor Programme of ESA. D. B. is thankful to Claudia Maraston for kindly providing the M13 models and useful comments on the manuscript, and to Rashmi Gottumukkala for support with the B24 spectra. D.B., P.S., R.D.P.,  F.L.B., A.G., A.P., C.S., and S.Z. acknowledge support by the grant PRIN-INAF-2019 1.05.01.85. F.L.B. acknowledges support from the INAF grant 1.05.23.04.01. This work made use of and acknowledge the following software: \texttt{NumPy} \citep{numpy}; \texttt{SciPy} \citep{scipy}; \texttt{Astropy} \citep{astropy}; \texttt{matplotlib} \citep{matplotlib}; \texttt{EAZY} \citep{eazy}; \texttt{pPXF} \citep{ppxf_2004, ppxf_2017, ppxf_2023}; \texttt{FAST} \citep{fast}; \texttt{emcee} \citep{emcee}. 
\end{acknowledgements}

\bibliographystyle{aa} % style aa.bst
\bibliography{biblio.bib} % your references Yourfile.bib

\begin{appendix}
\section{DR3 vs B24}\label{app:DR3_B24}
\subsection{Comparison of the observed spectra}\label{app:DR3_B24_spec}

\begin{figure}
\center
\includegraphics[width=0.92\columnwidth]{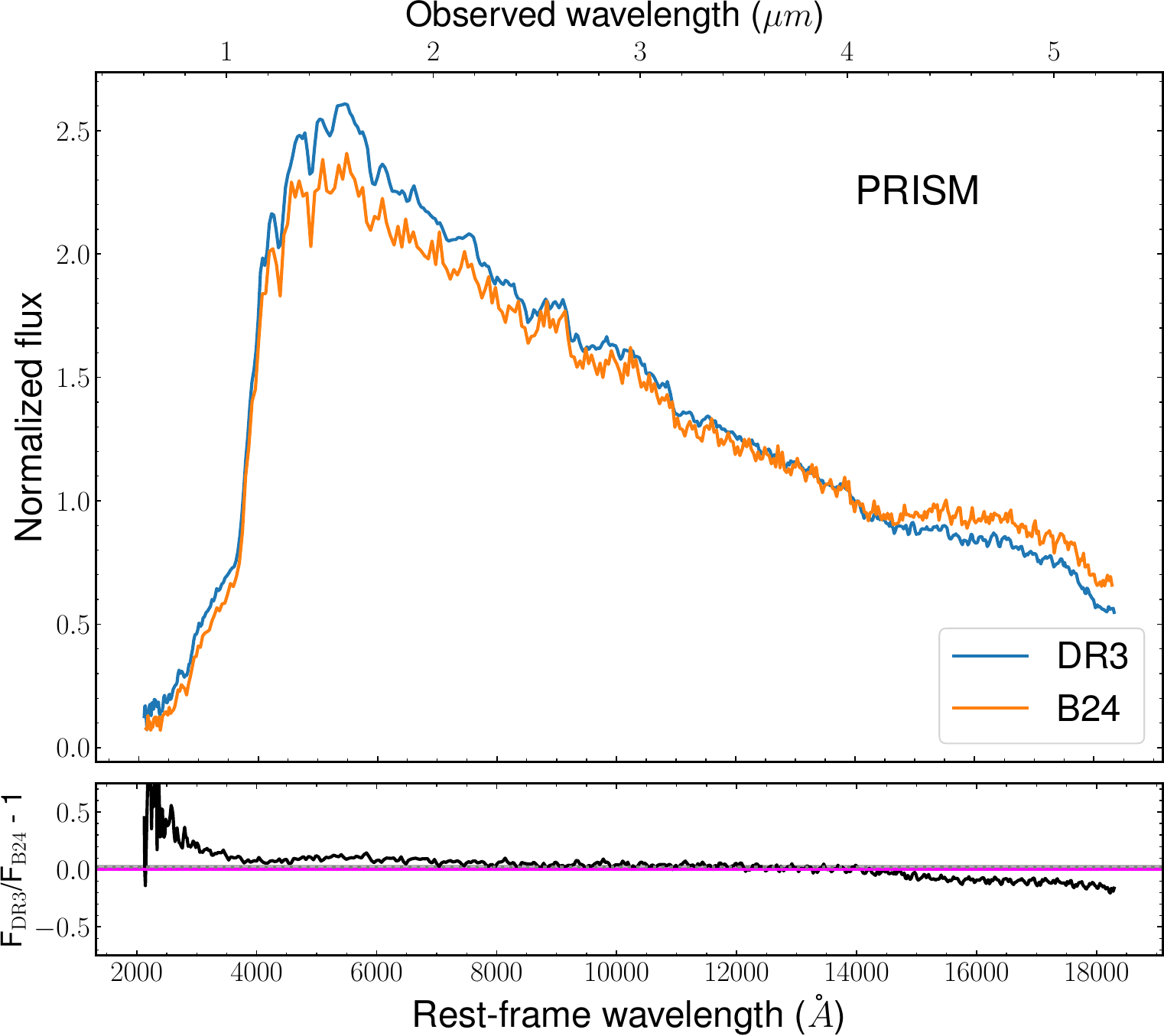}
\includegraphics[width=0.92\columnwidth]{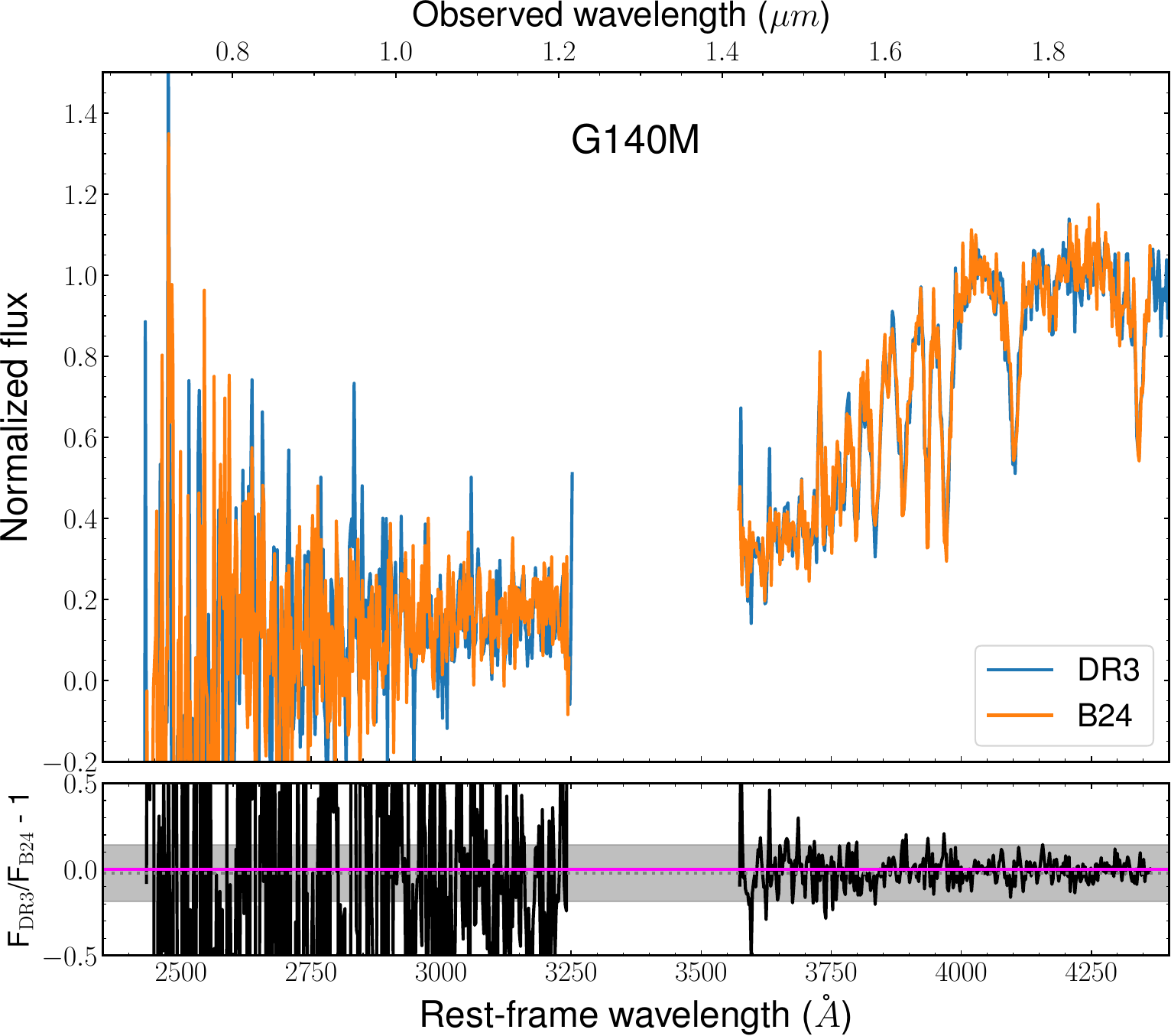}
\includegraphics[width=0.92\columnwidth]{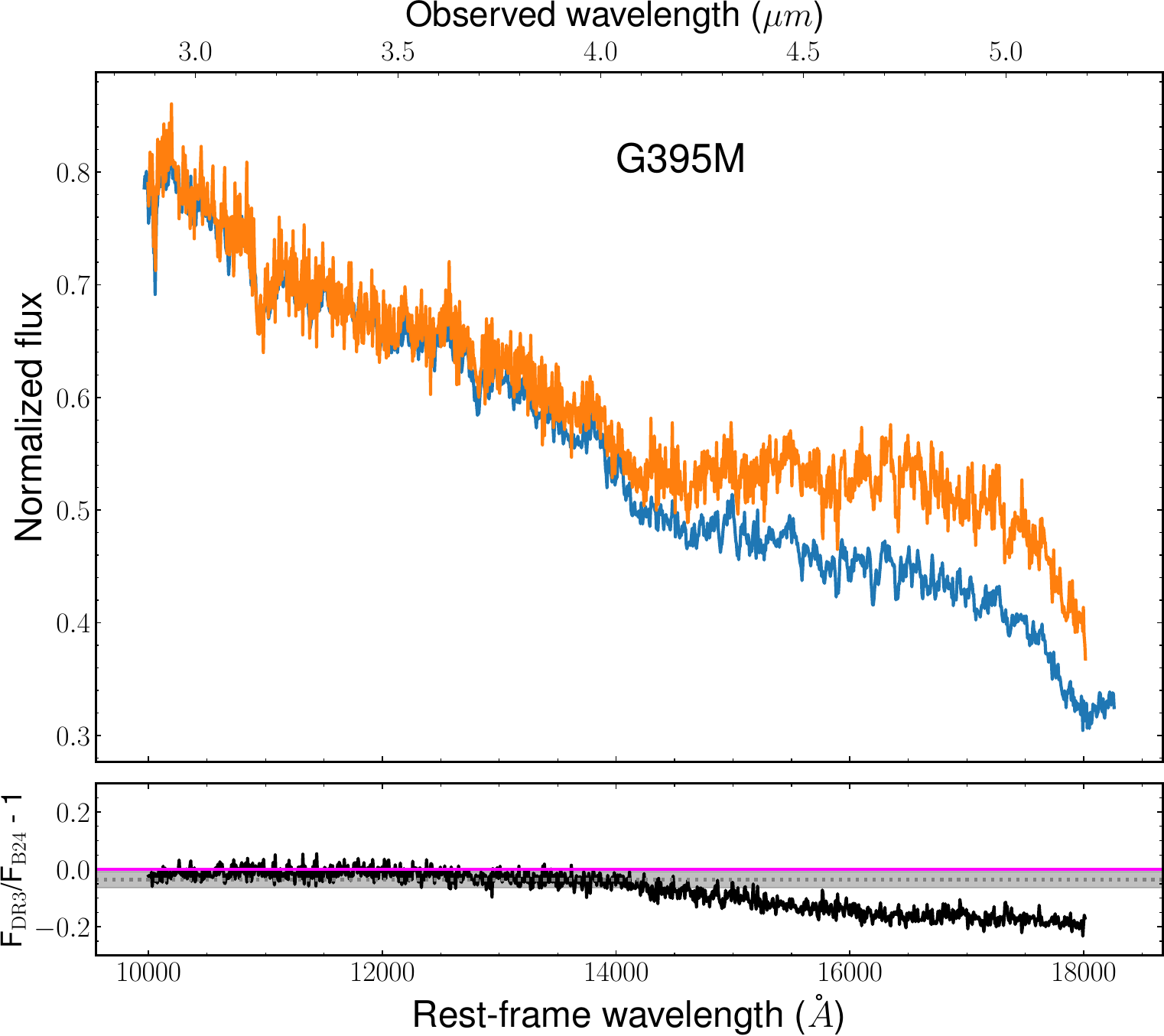}
\caption{Comparison of the G140M (upper panel), G395M (middle panel), and prism (lower panel) spectra for the data reduction of the JADES DR3 (orange) and B24 (blue). The black lines are the flux ratios F$_{\rm DR3}$/F$_{\rm B24} - 1$, the gray dotted lines and shaded regions are the median values and the quadratic errors on the flux, respectively, while the magenta line marks the zero.}
\label{fig:dr3_b24}
\end{figure}

In Fig. \ref{fig:dr3_b24} we show the comparison between the data reduction of the JADES DR3 and that of B24. In the upper panel we show the comparison of the prism spectra from the two reductions. In this case, there is a clear difference on the slope, varying from about a $+20 \%$ to $- 20\%$ between the two data reduction, with a peak of a $50\%$ excess in the DR3 prism spectrum at rest-frame wavelengths below $\sim 3000 \, \AA$. As shown in the next section, this difference leads to different estimates of the stellar population parameters.

\begin{figure}
\centering
\includegraphics[width=0.49\columnwidth]{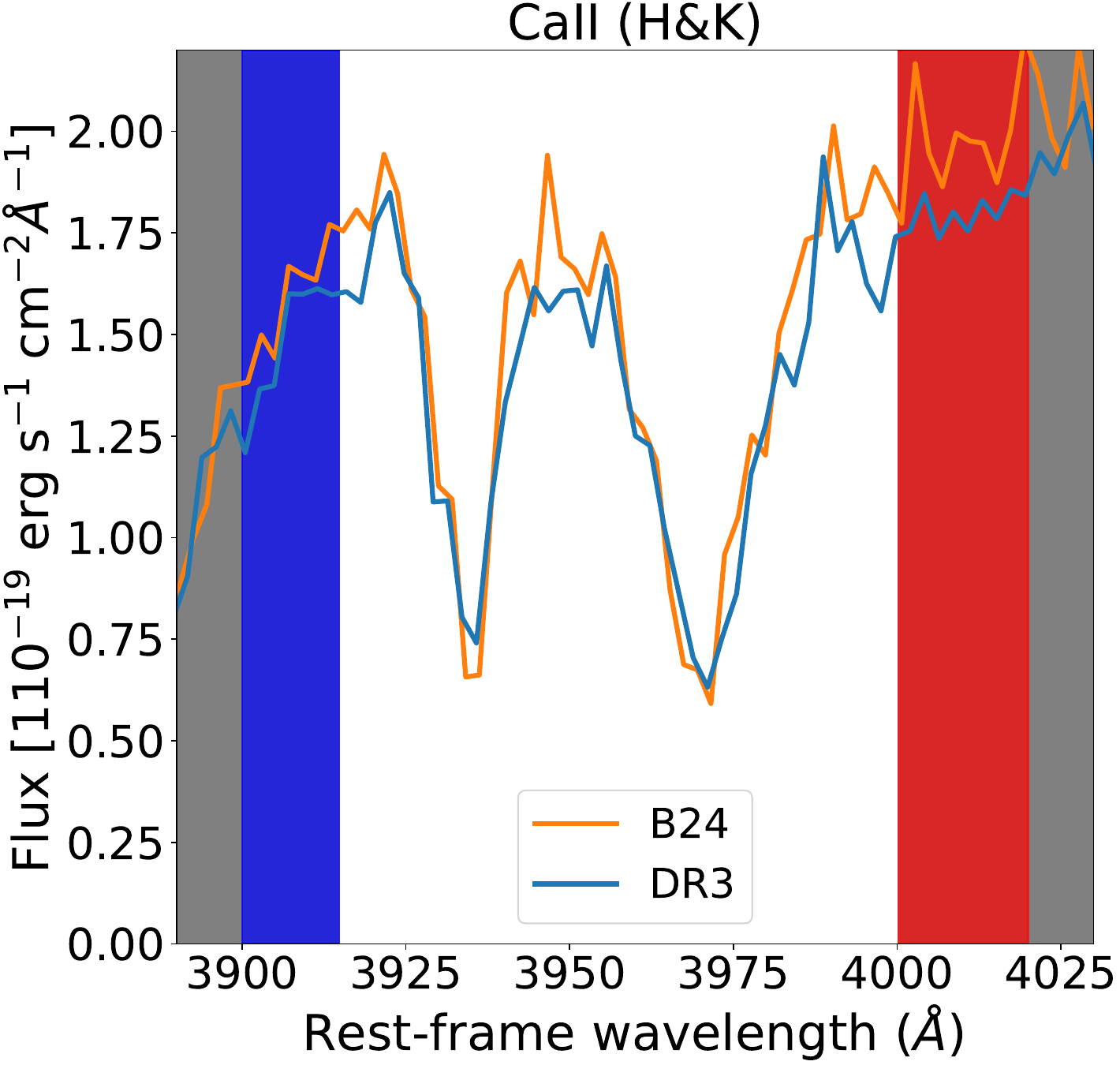}
\includegraphics[width=0.49\columnwidth]{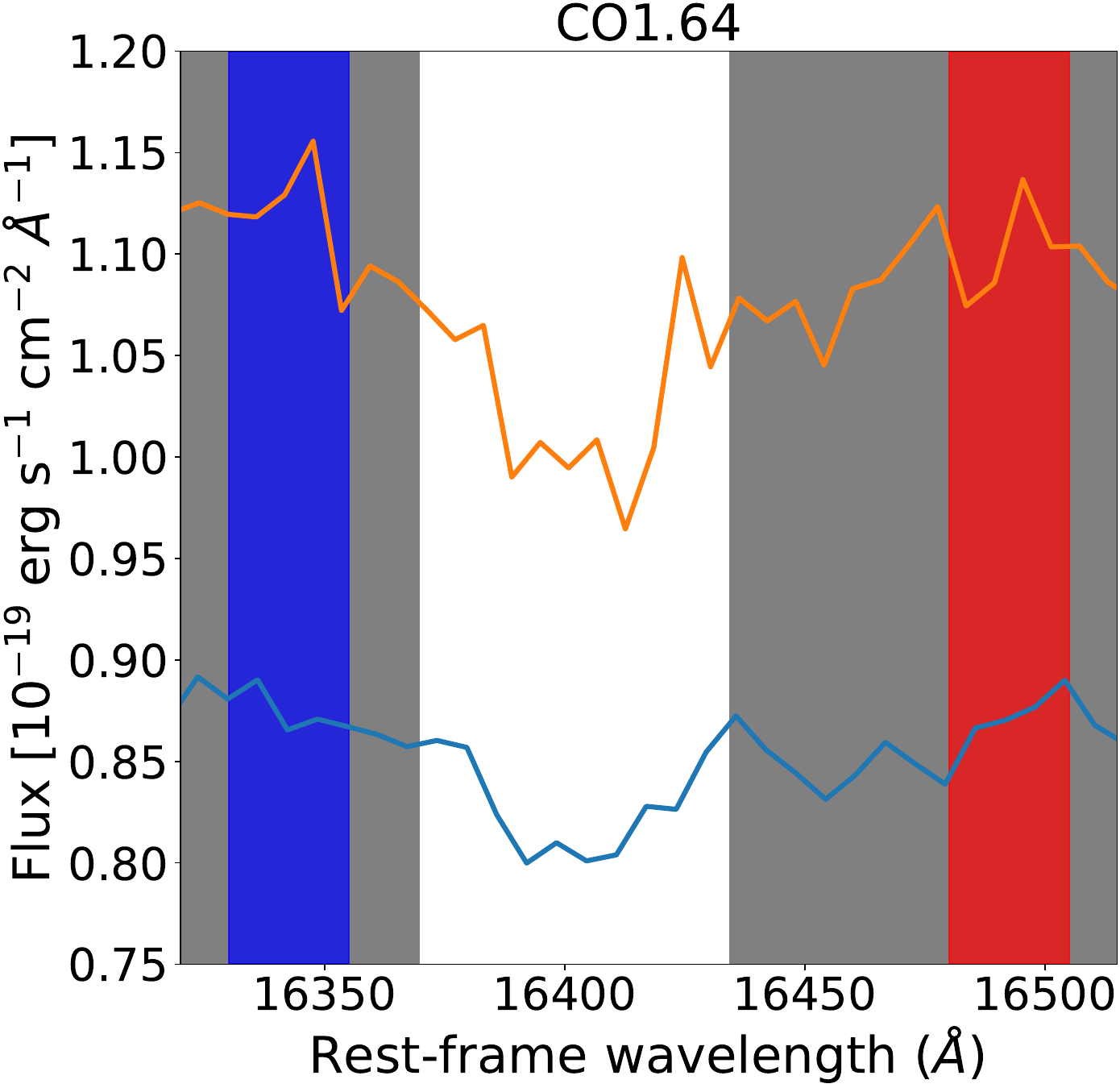}
\caption{CaII (H\&K) index (left panel) and CO1.64 index (right panel) for the two data reductions, DR3 (blue) and B24 (orange). The measurements of the former (latter) index are consistent (inconsistent) when measured from the two data reductions. In both panels, the white central region is the wavelength region where the feature is measured, while  blue and red regions are the blue and red continuum bandpasses, respectively.}
\label{fig:specind_dr3_b24}
\end{figure}

In the middle panel, we show the G140M. The two spectra are very similar, with a median difference in the flux close to zero, and they are consistent within the errors all over the spectrum, except for the NUV (observed $\lambda < 1.2 \mu m$), where the noise dominates. All the measured spectral indices in this wavelength range are also consistent when measured from the DR3 and B24. As an example example, in the upper panel of Fig. \ref{fig:specind_dr3_b24} we show the CaII-HK index.

Finally, in the lower panel, we show the G395M spectrum. In this case, the two spectra are consistent from $\sim 2.9 - 4 \mu m$ ($\sim 1 - 1.4 \mu m$ rest-frame), with a median systematic offset of only $\sim 5\%$. At longer wavelengths, however, the two spectra are clearly different, as the flux of the B24 spectrum is significantly higher than that of the DR3, increasing with the wavelength, and up to $\sim 20\%$ higher. We note that this difference is very similar to the one observed in the prism. Such differences do not affect only the continuum, but also the single absorption features. For example, in the lower panel of Fig. \ref{fig:specind_dr3_b24} we show the CO1.64 index of the two reductions. Here, there is not only a systematic offset in the flux, but the shape of the feature, as well as of the side bands is clearly different, leading to inconsistent measurements of the same index.

\subsection{Stellar population parameters}\label{app:DR3_B24_agemet}
\begin{figure}
\includegraphics[width=\columnwidth]{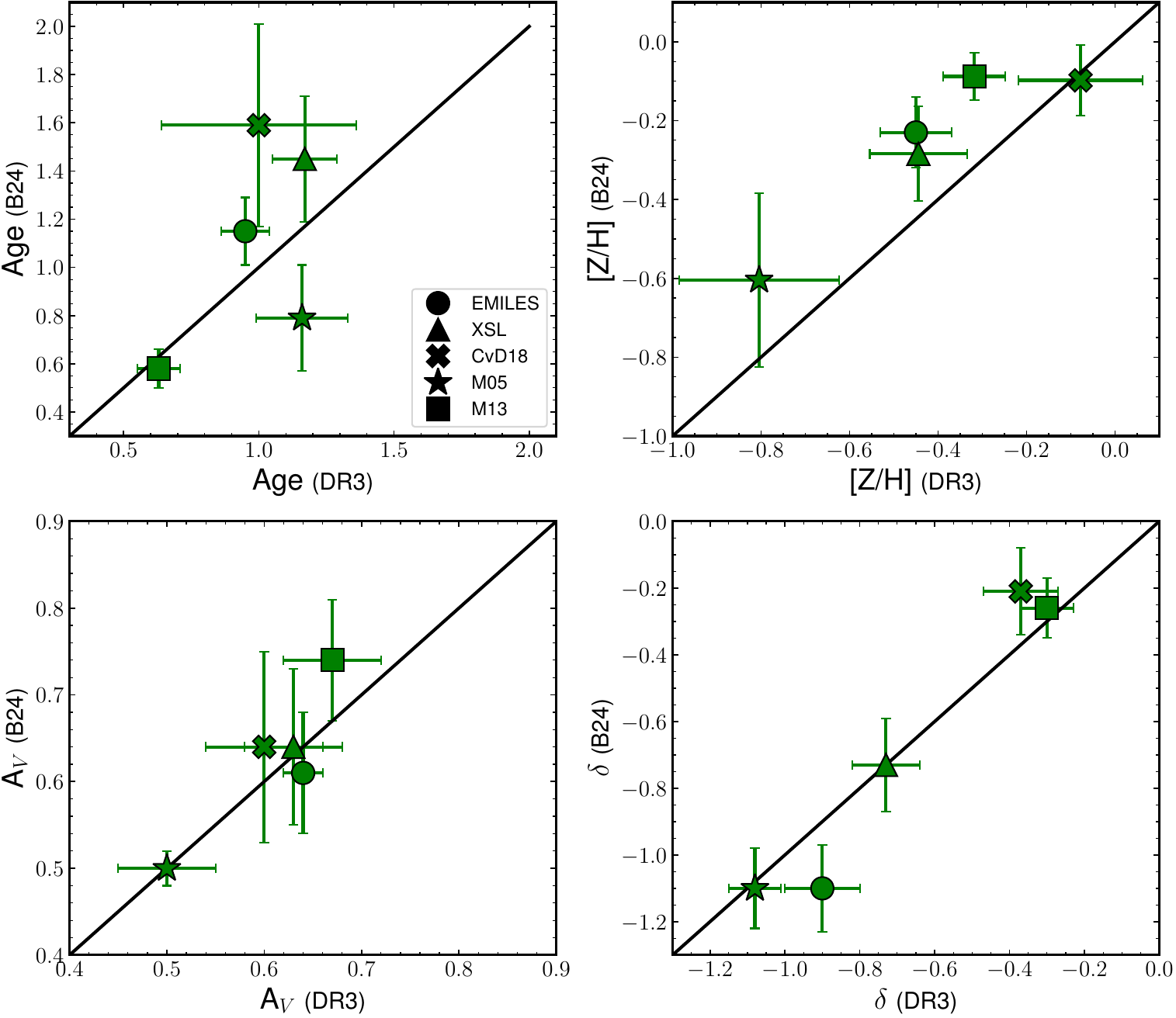}
\caption{Comparison of the different stellar population parameters estimated when fitting the DR3 and B24 prism spectra.}
\label{fig:agemet_dr3_b24}
\end{figure}

In Fig. \ref{fig:agemet_dr3_b24} we show the comparison between the stellar population parameters estimated for the DR3 and B24 prism spectra, using the same methods described in section \ref{sect:agemet} and different SSPs libraries (see also Appendix \ref{app:ssp}). The different slopes of the DR3 and B24 spectra generally provide different estimated stellar population parameters. Generally, the B24 spectrum provides systematically older ages (up to 0.5 Gyr) and higher metallicities (up to 0.2 dex). This is consistent with the B24 spectrum being flatter than the DR3 spectrum. The dust attenuation parameters, instead, are generally more consistent.

We verified that the G140M (fitted in the wavelength range $3600-4400 \, \AA$) and G395M (fitted in the wavelength range $1-1.4 \, \mu m$) provide consistent results for all parameters. This is not surprising, given the consistency of the two spectra in the wavelength regions considered (Fig. \ref{fig:dr3_b24}). 

\section{Stellar population parameters with different SSPs libraries}\label{app:ssp}

\begin{figure}
\centering
\includegraphics[width=0.7\columnwidth]{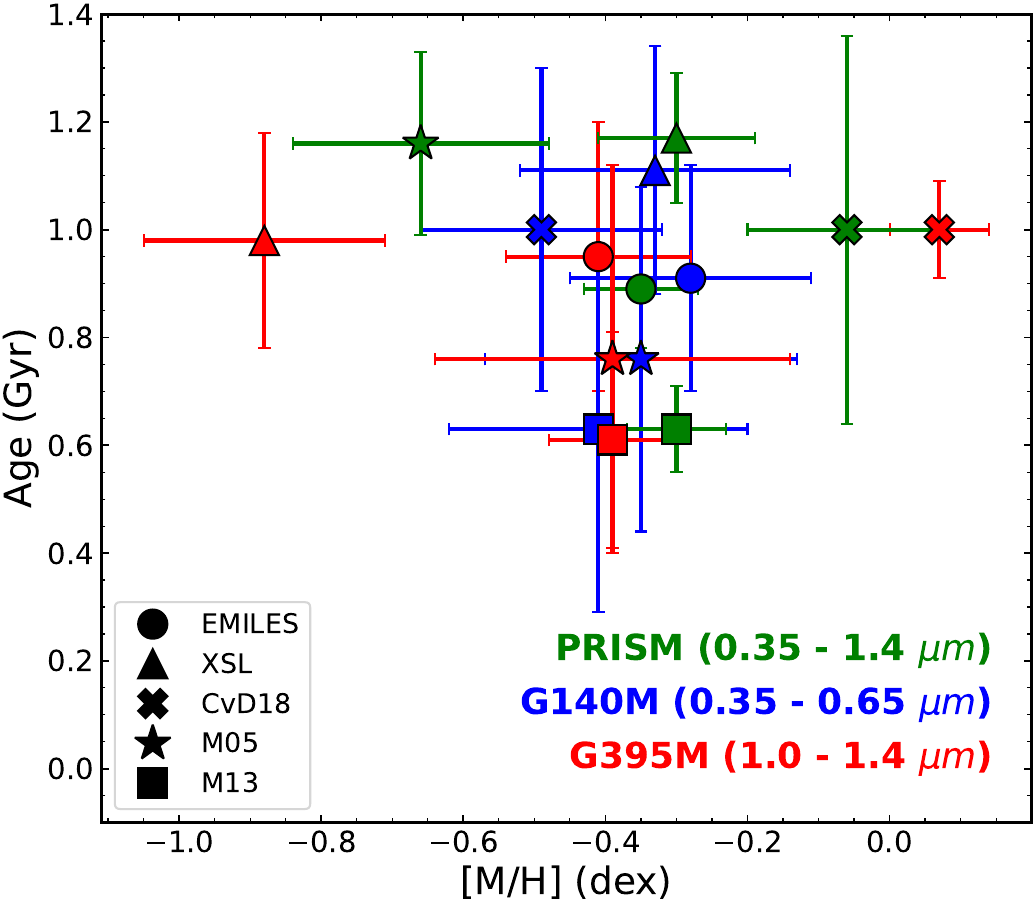}
\includegraphics[width=0.7\columnwidth]{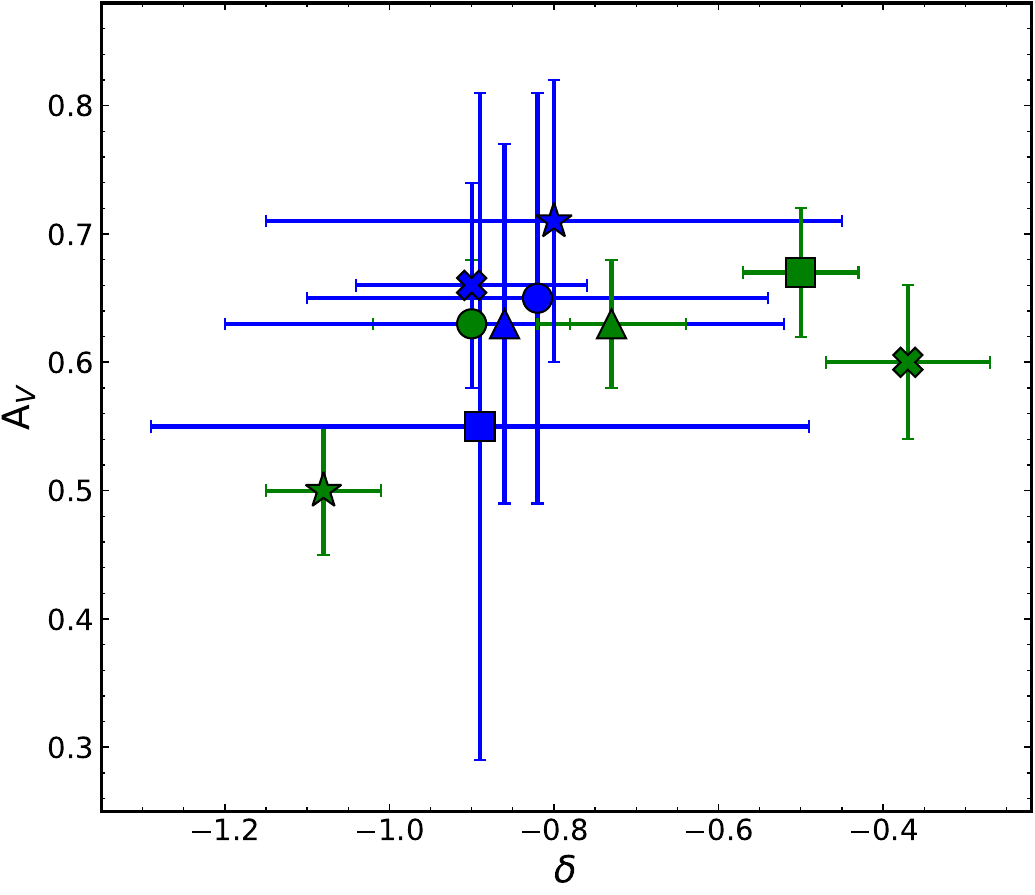}
\caption{Comparison of the stellar population parameters estimated using different SSP libraries}
\label{fig:agemet_SSPs}
\end{figure}

We fitted the stellar population parameters using the models libraries described in section \ref{sect:templates} (see also Table \ref{tab:libraries}). Since the M05 and M13 models have a lower resolution than the G140M and G395M spectra, we lower the resolution of the latter to match that of the models before performing the fit. We note that different models adopt different isochrones, having different solar metallicities, so we rescaled all the metallicities to $Z_\odot = 0.0198$ after the fit. In Fig. \ref{fig:agemet_SSPs} we show the results of the fits using different models libraries. In general, the estimates are consistent with those of the EMILES models. Considering all the different wavelength ranges and models libraries fitted we estimate averages age 0.94 Gyr, metallicity $-0.39$ dex, A$_V$ 0.65 mag, and $\delta$ -0.8 dex, consistent with those reported in \ref{sect:agemet}. We note that the M13 models estimate ages $\sim 0.6$ Gyr in all cases, that are the youngest values among the libraries. In section \ref{sect:cocn} and \ref{sect:indices} we show the M13 can match better many indices assuming a younger age for J-138717. We also note that the XSL models provide typically older ages. Finally, we note that the CvD18 models provide solar metallicity when fitting the prism and NIR spectrum, although this could be due to the lack of younger models; this could also explain the high value of $\delta$ measured from the prism.

\end{appendix}

\end{document}